\documentclass[preprint,jcp,aip]{revtex4-1}
\usepackage[latin1]{inputenc}
\usepackage{amsmath}
\usepackage{amssymb}
\usepackage{graphicx}

\makeatletter

\providecommand{\tabularnewline}{\\}

\usepackage{appendix}
\usepackage{xcolor}
\usepackage{nicefrac}
\usepackage{multirow}
\usepackage{tikz}
\usetikzlibrary{shapes.geometric}
\usetikzlibrary{decorations.pathmorphing}
\usetikzlibrary{patterns}

\newcommand{\sectionabb}{\text{Sec.\ }}
\newcommand{\appendixabb}{\text{App.\ }}
\newcommand{\equationabb}{\text{Eq.\ }}

\newcommand{\equationsabb}{\text{Eqs.\ }}

\newcommand{\figureabb}{\text{Fig.\ }}

\newcommand{\tableabb}{\text{Tab.\ }}

\makeatother

\begin{document}

\title{Simplified Approach to the Mixed Time-averaging Semiclassical Initial
Value Representation for the Calculation of Dense Vibrational Spectra}

\author{Max \surname{Buchholz}}

\affiliation{Dipartimento di Chimica, Università degli Studi di Milano, via C.
Golgi 19, 20133 Milano, Italy}

\author{Frank \surname{Grossmann}}

\affiliation{Institut für Theoretische Physik, Technische Universität Dresden,
01062 Dresden, Germany }

\author{Michele \surname{Ceotto}}

\affiliation{Dipartimento di Chimica, Università degli Studi di Milano, via C.
Golgi 19, 20133 Milano, Italy}
\email{michele.ceotto@unimi.it}

\begin{abstract}
We present and test an approximate method for the semiclassical calculation
of vibrational spectra. The approach is based on the mixed time-averaging
semiclassical initial value representation method, which is simplified
to a form that contains a filter to remove contributions from approximately
harmonic environmental degrees of freedom. This filter comes at no
additional numerical cost, and it has no negative effect on the accuracy
of peaks from the anharmonic system of interest. The method is successfully
tested for a model Hamiltonian, and then applied to the study of the
frequency shift of iodine in a krypton matrix. Using a hierarchic
model with up to 108 normal modes included in the calculation, we
show how the dynamical interaction between iodine and krypton yields
results for the lowest excited iodine peaks that reproduce experimental
findings to a high degree of accuracy.
\end{abstract}
\maketitle

\section{Introduction}

Since the early seventies of the past century, quantum molecular dynamics
has been devoted to the study of gas phase reactions on pre-computed
potential energy surfaces.\cite{miller1968uniform,wolken1972theoretical,garrett1978quantum,miller1971semiclassical,bowman1980sudden,bowman1973semi,schatz1975exact,gianturco1976scattering,aquilanti1976total,aquilanti1975resonant,bonnet1997quasiclassical,aquilanti2000collective,connor1979comparison,de1994semiclassical,Miller_Ceotto_2003QI,Ceotto_Miller_2004test,Ceotto_Yang_2005quantum,Ceotto_Aieta2017quantumTST,Ceotto_Aieta_Gabas_16,Ceotto_Mandra2013deep,Ceotto_Mandra2014helium,Conte_Bowman_CollisionsCH4-H2O_2015,Conte_Bowman_Manybody_2015,Homayoon_Bowman_H2-H2O_2015,Houston_Bowman_RoamingH2CO_2016,Chen_Bowman_Methane-Water_2015,Conte_Bowman_Communication_2014,Conte_Houston_Bowman_Ar2013}
However, condensed phase nuclear quantum molecular dynamics has gradually
attracted more and more attention from researchers mostly for its
practical applications. The question if quantum mechanical effects
are important and crucial for the description of nuclear condensed
phase phenomena is still open. Most probably the answer would be:
``\emph{it depends}''. Spectroscopy shows that nuclear energy levels
are quantized even if the full dimensional spectrum could appear as
a continuum. 

Several approaches to condensed phase dynamics are based on path integrals
(PI).\cite{feynman_pathintegral_1965} In methods such as PI Monte
Carlo (PIMC) \cite{Ceperley1994_reviewPIMC} and PI molecular dynamics
(PIMD), \cite{parrinello1984study,marx1996ab,tuckerman1996efficient}
thermodynamic properties are calculated by considering the imaginary
time propagator for the Boltzmann operator. More recently, also real-time
dynamics studies based on path integrals have been performed. There
exist several methods, such as centroid path integral molecular dynamics
(CPMD) \cite{cao1994formulation,geva2001quantum,paesani2009infrared,poulsen2001path}
and ring polymer molecular dynamics (RPMD).\cite{habershon2013ring,craig2005chemical,craig2005refined,habershon2007quantum,markland2008efficient,richardson2009ring,suleimanov2011bimolecular,menzeleev2011direct,ananth2013mapping,shakib2017ring,witt2009applicability,rossi_manolopoulos_TRPDM_2014}
There, the dynamics of the nuclei is treated quantum mechanically
by mapping them onto fictitious classical particles connected by springs.
A critical review of those methods with respect to their applicability
to vibrational spectroscopy can be found in Ref.$\ $\onlinecite{witt2009applicability}.

Also in semiclassical molecular dynamics, real as well as imaginary
time propagations can be performed.\cite{miller1971semiclassical,Miller2006,Heller_TdependentSC_1975,Kay_Atomsandmolecules_2005,Thoss_Wang_SemiclassicalReview_2004,Walton_Manolopoulos_FranckCondon_1996,Walton_Manolopoulos_FrozenGaussianCO2_1995,Pollak_Perturbationseries_2007,Bonella_Coker_Linearizedpathintegral_2005,Huo_Coker_Semiclassicalnonadiabatic_2012,Miller_Addingquantumtoclassical_2001,Miller_PNAScomplexsystems_2005,Kay_Multidim_1994,Kay_Numerical_1994,Heller_FrozenGaussian_1981,Wang_Miller_Chemicalreactions_1998,Yamamoto_Miller_Fluxcorrelation_2002,Conte_Pollak_ContinuumLimit_2012,Conte_Pollak_ThawedGaussian_2010}
These methods can also be derived from path integrals\cite{Berry_Mount_Semiclassical_1972}
and they have been applied both to gas phase problems\cite{Wang2000,Wang2001,Zhuang_Ceotto_Hessianapprox_2012,ceotto_conte_DCSCIVR_2017,Gabas_Ceotto_Glycine_2017,Monteferrante_Ciccotti_Liquidneon_2013,Bonella_Coker_Linearizedpathintegral_2005,Bonella_Kapral_quantum-classical_2010,Pollak_MFierro_FBIVR_2007,Wehrle_Vanicek_NH3_2015,Wehrle_Vanicek_Oligothiophenes_2014,Ceotto_AspuruGuzik_Multiplecoherent_2009,Ceotto_AspuruGuzik_PCCPFirstprinciples_2009,Ceotto_Tantardini_Copper100_2010,Ceotto_AspuruGuzik_Curseofdimensionality_2011,Ceotto_AspuruGuzik_Firstprinciples_2011,Tamascelli_Ceotto_GPU_2014,Conte_Ceotto_NH3_2013,Ceotto_Hase_AcceleratedSC_2013,Tatchen_Pollak_Onthefly_2009,Wong_Roy_Formaldehyde_2011}
and to model potentials of condensed phase systems, such as the Caldeira-Leggett
potential.\cite{Wang_Miller_Chemicalreactions_1998,thoss2001self,Ceotto_Buchholz_MixedSC_2017}
This paper deals with the application of semiclassical initial value
representation (SCIVR)\cite{miller1971semiclassical,Miller_PNAScomplexsystems_2005,miller1998spiers,miller2001semiclassical,Heller_Cellulardynamics_1991,Heller_FrozenGaussian_1981,Heller_SCspectroscopy_1981,Herman1994,Herman_Kluk_SCnonspreading_1984,Kay_Integralexpression_1994,Kay_Multidim_1994,Kay_Numerical_1994,Kay_SCcorrections_2006,Pollak_Perturbationseries_2007,Pollak_MFierro_FBIVR_2007,MFierro_Pollak_FBIVR_2006,Grossmann1995,Grossmann_HierarchySC_1999,Sun1997,Sun1998,Sun1998-1,Miller_S-Matrix_1970,Heller_TdependentSC_1975,Grossmann_Xavier_SCderivation_1998}
molecular dynamics to condensed phase systems. More specifically,
we recently designed a SCIVR method called mixed time-averaging SCIVR
(M-TA-SCIVR)\cite{Buchholz_Ceotto_MixedSC_2016} for the calculation
of nuclear spectra for condensed phase systems composed of a main
system of interest (SOI) coupled to a bath. It employs the hybrid
dynamics idea\cite{Grossmann_SChybrid_2006} and is designed for SCIVR
nuclear power spectra calculations from the Fourier transform of a
wavepacket's correlation functions. In M-TA-SCIVR the environment
is treated by integrating out the phase space coordinates for the
corresponding degrees of freedom using a thawed Gaussian approximation.\cite{Heller_TdependentSC_1975}
The method is applicable to both pre-computed and on-the-fly ab initio
quantum dynamics simulations and it is free of any adjustable parameters.
M-TA-SCIVR proved to be reliable when compared to exact quantum results
for small dimensional systems.\cite{Buchholz_Ceotto_MixedSC_2016}
Furthermore, in an application to an anharmonic SOI coupled to a Caldeira-Leggett
environment with up to 60 harmonic bath degrees of freedom, good agreement
was found with respect to higher-accuracy SCIVRs.\cite{Ceotto_Buchholz_MixedSC_2017}

In this paper, we focus on the application of M-TA-SCIVR to problems
where both system and bath are anharmonic. This is quite challenging
due to the presence of (many) bath overtones in the spectrum, which
complicate peak attribution or render it altogether impossible. One
way to resolve this issue would be to start from initial conditions
where the bath modes have little or no initial energy. However, this
introduces a sampling bias because the classical dynamics explores
only the low energy, harmonic regions of the respective bath sites.
We will therefore introduce a simplified approach to M-TA-SCIVR (SAM-TA-SCIVR)
which acts as a filter for the bath excitations while still reproducing
exact system frequencies. We will apply SAM-TA-SCIVR to the power
spectrum of an iodine molecule in a krypton matrix, since this is
a well studied complex condensed phase system.\cite{Coker_2008I2Kr,Ovchinnikov_Apkarian_Condensedphase_1996,Ovchinnikov_Apkarian_Ramanspectra_1997}
It is realized that the full dimensional spectrum is very dense and
that a technique, which is able to decompose the spectrum into specific
components pertaining to the normal modes of interest, would be very
useful for the interpretation and for a better understanding of the
physics. For these reasons, we describe how to selectively extract
the spectrum of the SOI without resorting to any artificial decoupling
from the environment.

The paper is organized in the following way: Sec.$\ $\ref{sec:mixed}
recalls the M-TA-SCIVR method (\ref{subsec:Mixed-Time-averaging-Semiclassic})
and presents the new approximation for dense spectra calculations
(\ref{subsec:SAM_derivation}). In Sec.$\ $\ref{sec:results} some
tests on model systems are reported followed by the main application
which is the calculation of the power spectra for the iodine molecule
in a krypton matrix. Conclusions are drawn and future perspectives
are given in Sec.$\ $\ref{sec:conc}.

\section{Simplified Approach to the Mixed Time-averaging Semiclassical Initial
Value Representation}

\label{sec:mixed}

The main idea of this paper is to propose a method for the calculation
of molecular spectra that has a built-in filter, removing unwanted
contributions from environmental degrees of freedom (DOFs). The need
for such a filter arises, when the spectrum becomes too noisy for
unambiguous peak identification, which may be the case if many DOFs
carry initial excitation. As this approach is a simplification of
the recently introduced M-TA-SCIVR,\cite{Buchholz_Ceotto_MixedSC_2016,Ceotto_Buchholz_MixedSC_2017}
we first give a brief overview of its derivation and then continue
with a simplification that allows for the treatment of systems with
possibly hundreds of degrees of freedom.

\subsection{Mixed Time-averaging Semiclassical Initial Value Representation\label{subsec:Mixed-Time-averaging-Semiclassic}}

The quantity to be calculated with M-TA-SCIVR is the power spectrum
$I(E)$ of a given initial state $\left|\chi\right\rangle $ subject
to a Hamiltonian $\hat{H.}$ It can be found from the system's dynamics
as the Fourier transform of the autocorrelation function
\begin{align}
I(E)=\frac{1}{2\pi\hbar}\int\limits _{-\infty}^{\infty}\text{d}t\ \text{e}^{\text{i}Et/\hbar}\left\langle \chi\middle|\text{e}^{-\text{i}\hat{H}t/\hbar}\middle|\chi\right\rangle .\label{eq:IE_dyn}
\end{align}
The time evolution in \equationabb (\ref{eq:IE_dyn}) is calculated
semiclassically with the propagator of Herman and Kluk \cite{Herman_Kluk_SCnonspreading_1984},
\begin{align}
\text{e}^{-\text{i}\hat{H}t/\hbar}=\  & \frac{1}{(2\pi\hbar)^{F}}\int\text{d}\mathbf{p}(0)\int\text{d}\mathbf{q}(0)\ C_{t}(\mathbf{p}(0),\mathbf{q}(0))\nonumber \\
 & \quad\times\text{e}^{\text{i}S_{t}(\mathbf{p}(0),\mathbf{q}(0))/\hbar}\left|\mathbf{p}(t),\mathbf{q}(t)\right\rangle \left\langle \mathbf{p}(0),\mathbf{q}(0)\right|,\label{eq:HK_propagator}
\end{align}
where $(\mathbf{p}(t),\mathbf{q}(t))$ is the $2F$-dimensional classical
phase space trajectory evolving from initial conditions $(\mathbf{p}(0),\mathbf{q}(0))$,
and $S_{t}$ is the corresponding classical action. \equationabb
(\ref{eq:HK_propagator}) also contains the HK prefactor, 
\begin{align}
C_{t} & (\mathbf{p}(0),\mathbf{q}(0))=\nonumber \\
 & \sqrt{\frac{1}{2^{F}}\text{det}\left[\frac{\partial\mathbf{q}(t)}{\partial\mathbf{q}(0)}+\frac{\partial\mathbf{p}(t)}{\partial\mathbf{p}(0)}-\text{i}\hbar\boldsymbol{\gamma}\frac{\partial\mathbf{q}(t)}{\partial\mathbf{p}(0)}+\frac{\text{i}}{\hbar\boldsymbol{\gamma}}\frac{\partial\mathbf{p}(t)}{\partial\mathbf{q}(0)}\right]}\label{eq:HK_prefactor}
\end{align}
which accounts for second-order quantum delocalizations around the
classical paths. Finally, the coherent state basis set in position
representation for many degrees of freedom is given by the direct
product of one-dimensional Gaussian wavepackets, 
\begin{align}
\left\langle \mathbf{x}|\mathbf{p},\mathbf{q}\right\rangle  & =\left(\frac{\det(\boldsymbol{\gamma})}{\pi^{F}}\right)^{1/4}\nonumber \\
 & \times\exp\left[-\frac{1}{2}\left(\mathbf{x}-\mathbf{q}\right){}^{\text{T}}\boldsymbol{\gamma}\left(\mathbf{x}-\mathbf{q}\right)+\frac{\text{i}}{\hbar}\mathbf{p}^{\text{T}}\left(\mathbf{x}-\mathbf{q}\right)\right]\label{eq:HK_gaussians}
\end{align}
where $\boldsymbol{\gamma}$ is a diagonal matrix containing $F$
time independent width parameters.

While the semiclassical approximation of the propagator in \equationabb
(\ref{eq:HK_propagator}) in principle allows for the inclusion of
an arbitrary number of DOFs, practical applications are limited by
the need to converge the phase space integral. We will therefore carry
out two steps to accelerate the numerical Monte Carlo phase space
integration of \equationabb (\ref{eq:HK_propagator}). The first
step is the introduction of a time averaging integral, \cite{Kaledin_Miller_Timeaveraging_2003,Elran_Kay_ImprovingHK_1999}
which is applied to \equationabb (\ref{eq:IE_dyn}) and yields a
semiclassical approximation with a pre-averaged phase space integrand.
This expression can be further simplified with Kaledin and Miller's
so-called separable approximation \cite{Kaledin_Miller_TAmolecules_2003}
that results in 
\begin{align}
 & I(E)=\frac{1}{\left(2\pi\hbar\right)^{F}}\frac{1}{2\pi\hbar T}\int\text{d}\mathbf{p}(0)\int\text{d}\mathbf{q}(0)\nonumber \\
 & \times\left|\int\limits _{0}^{T}\text{d}t\left\langle \chi\middle|\mathbf{p}(t),\mathbf{q}(t)\right\rangle \text{e}^{\text{i}\left[S_{t}\left(\mathbf{p}(0),\mathbf{q}(0)\right)+Et+\phi_{t}\left(\mathbf{p}(0),\mathbf{q}(0)\right)\right]/\hbar}\right|^{2},\label{eq:HK_sep}
\end{align}
where $\phi_{t}\left(\mathbf{p}(0),\mathbf{q}(0)\right)$ denotes
the phase of the HK prefactor $C_{t}\left(\mathbf{p}(0),\mathbf{q}(0)\right).$
The expression now contains a positive-definite phase space integrand.
While less computationally demanding than \equationabb (\ref{eq:HK_propagator}),
the separable approximation TA-SCIVR in \equationabb (\ref{eq:HK_sep})
has also turned out to be very accurate for a number of molecular
dynamics applications. \cite{Kaledin_Miller_TAmolecules_2003,Kaledin_Miller_Timeaveraging_2003,Ceotto_AspuruGuzik_Curseofdimensionality_2011,Ceotto_AspuruGuzik_Firstprinciples_2011,Ceotto_AspuruGuzik_Multiplecoherent_2009,Ceotto_AspuruGuzik_PCCPFirstprinciples_2009,Ceotto_Buchholz_MixedSC_2017,Buchholz_Ceotto_MixedSC_2016,DiLiberto_Ceotto_Prefactors_2016,Gabas_Ceotto_Glycine_2017,Tamascelli_Ceotto_GPU_2014,Conte_Ceotto_NH3_2013,Ceotto_Hase_AcceleratedSC_2013,Zhuang_Ceotto_Hessianapprox_2012}

The second step towards making the dynamics of larger systems accessible
is to invoke the mixed approximation. To this end, we use the semiclassical
hybrid dynamics idea \cite{Grossmann_SChybrid_2006} to divide the
$2F$ phase space variables into $2F_{\text{hk}}$ for the system
space and $2F_{\text{tg}}$ for the bath phase space. Only the system
part, denoted by the subscript \emph{hk}, is then treated on the HK
level of accuracy, whereas the simpler single-trajectory TGWD approximation
is used for the bath DOFs, which are denoted by the subscript \emph{tg}.
This separation is made only for the semiclassical expression, while
the underlying classical dynamics is not modified. We now assume a
reference state of Gaussian form, $\left|\chi\right\rangle =\left|\mathbf{p}_{\text{eq}},\mathbf{q}_{\text{eq}}\right\rangle $,
where $\mathbf{q_{\text{eq}}}$ is the equilibrium position and $\mathbf{p}_{\text{eq}}$
is the momentum corresponding to some approximated eigenenergy. In
the mixed approximation, the initial phase space coordinates $\left(\mathbf{p}(0),\mathbf{q}(0)\right)$
are redefined as 
\begin{align}
\mathbf{p}(0)=\left(\begin{matrix}\mathbf{p}_{\text{hk}}(0)\\
\mathbf{p}_{\text{eq},\text{tg}}
\end{matrix}\right),\qquad\mathbf{q}(0)=\left(\begin{matrix}\mathbf{q}_{\text{hk}}(0)\\
\mathbf{q}_{\text{eq},\text{\text{tg}}}
\end{matrix}\right).\label{eq:HYB_initial}
\end{align}
 Only the HK initial conditions $\left(\mathbf{p}_{\text{hk}}(0),\mathbf{q}_{\text{hk}}(0)\right)$
are found by Monte Carlo sampling around $\left(\mathbf{p}_{\text{eq},\text{\text{hk}}},\mathbf{q}_{\text{eq},\text{\text{hk}}}\right)$,
while the bath starting coordinates are always at the equilibrium
positions, $\left(\mathbf{p}_{\text{tg}}(0),\mathbf{q}_{\text{tg}}(0)\right)=\left(\mathbf{p}_{\text{eq},\text{tg}},\mathbf{q}_{\text{eq},\text{\text{tg}}}\right)$.
Since the TGWD is exact for harmonic potentials, this division should
accurately reproduce the contributions of weakly coupled bath DOFs
close to their potential minimum. With this phase space division in
place, we expand the classical trajectories and the action to first
and second order, respectively, in the displacement coordinates of
the bath subspace. This approximates the exponent in \equationabb
(\ref{eq:HK_sep}) such that the phase space integration over the
original bath initial conditions $(\mathbf{p}_{\text{tg}}(0),\mathbf{q}_{\text{tg}}(0))$
can be performed analytically as a Gaussian integral, and the dimensionality
of the phase space integration is reduced. The resulting twofold time
integration collapses into a single one after another separable approximation
assuming approximately harmonic behavior of the bath, and we arrive
at the separable mixed TA-SCIVR (M-TA-SCIVR)
\begin{align}
I(E)= & \ \frac{1}{\left(2\hbar\right)^{F}}\frac{1}{\pi^{F_{\text{hk}}}}\frac{1}{2\pi\hbar T}\int\text{d}\mathbf{p}_{\text{hk}}\left(0\right)\int\text{d}\mathbf{q}_{\text{hk}}\left(0\right)\left|\int_{0}^{T}\text{d}t\:\text{e}^{\text{i}\left[Et+\phi_{t}\left(\mathbf{p}\left(0\right),\mathbf{q}\left(0\right)\right)+S_{t}\left(\mathbf{p}\left(0\right),\mathbf{q}\left(0\right)\right)\right]/\hbar}\right.\nonumber \\
 & \times\left\langle \mathbf{p}_{\text{eq},\text{\text{hk}}},\mathbf{q}_{\text{eq},\text{\text{hk}}}\middle|\mathbf{p}_{\text{\text{hk}}}\left(t\right),\mathbf{q}_{\text{\text{hk}}}\left(t\right)\right\rangle \left\langle \mathbf{p}_{\text{eq},\text{tg}},\mathbf{q}_{\text{eq},\text{\text{tg}}}\middle|\mathbf{p}_{\text{tg}}\left(t\right),\mathbf{q}_{\text{\text{tg}}}\left(t\right)\right\rangle \nonumber \\
 & \times\left.\frac{1}{\left[\det\left(\mathbf{A}\left(t\right)+\mathbf{A}^{*}\left(t\right)\right)\right]^{1/4}}\exp\left\{ \frac{1}{4}\mathbf{b}_{t}^{\text{T}}\left(\mathbf{A}\left(t\right)+\mathbf{A}^{*}\left(t\right)\right)^{-1}\mathbf{b}_{t}\right\} \right|^{2}.\label{eq:HYB_sep}
\end{align}
The matrix $\mathbf{A}(t)$ and the vector $\mathbf{b}(t)$ are defined
in App.$\ $\ref{sec:mixed_quantities}, and their contributions will
turn out to vanish with the simplification in Sec.$\ $\ref{subsec:SAM_derivation}.
As it has been demonstrated for a Morse oscillator embedded in a Caldeira-Leggett
bath with up to 61 DOFs, \cite{Buchholz_Ceotto_MixedSC_2016,Ceotto_Buchholz_MixedSC_2017}
M-TA-SCIVR reproduces both system and bath peaks precisely when compared
to exact quantum dynamics and full HK TA-SCIVR results, and reaches
tight convergence within a considerably shorter amount of time than
the separable TA-SCIVR from \equationabb (\ref{eq:HK_sep}).\cite{Buchholz_Ceotto_MixedSC_2016}

\subsection{Simplification and Bath Frequency Filter\label{subsec:SAM_derivation}}

Regarding the applicability of \equationabb(\ref{eq:HYB_sep}) to
large molecular systems, both time-averaging and phase space separation
put forward the convergence of the phase space integration with fewer
trajectories. However, one major drawback is not addressed: When investigating
a system with more than a handful of coupled degrees of freedom, spectra
from both TA-SCIVR and M-TA-SCIVR become very noisy if all degrees
of freedom carry some initial excitation. Contributions from excited
peaks, whose number grows exponentially with system size, make it
impossible to identify specific excitations even on the single-trajectory
level. Due to the positive definite nature of the phase space integrand
in \equationabb(\ref{eq:HK_sep}) and (\ref{eq:HYB_sep}), the phase
space average does not resolve this issue. An elegant solution has
been proposed in the form of multiple coherent states TA-SCIVR (MC-TA-SCIVR),
\cite{Ceotto_AspuruGuzik_Curseofdimensionality_2011,Ceotto_AspuruGuzik_Multiplecoherent_2009}
where the usual product reference state $\left|\chi\right\rangle $
in \equationabb(\ref{eq:HK_sep}) is replaced with a superposition
of states. This approach needs only a handful of trajectories with
initial conditions $\left(\mathbf{p}_{\text{eq}}^{i},\mathbf{q}_{\text{eq}}^{i}\right)$
chosen such that the classical energies are close to the positions
of the desired peaks in order to reproduce quantum results with high
accuracy. More importantly here, however, is that the reference state
in the MC TA-SCIVR approach can also be used as a filter. Choosing
the reference state, for example, as 
\begin{align}
\left|\chi\right\rangle = & \prod_{j=1}^{F}\left(\left|+p_{\text{eq},\:j},\:q_{\text{eq},\:j}\right\rangle +\left|-p_{\text{eq},\:j},\:q_{\text{eq},\:j}\right\rangle \right),\label{eq:filter_MCTASCIVR}
\end{align}
all odd contributions from the single-trajectory spectrum are removed,
thus reintroducing clearly distinguishable peaks.\cite{Ceotto_AspuruGuzik_Curseofdimensionality_2011}
This can be shown analytically for the harmonic oscillator, and also
works very well for anharmonic systems. The size of the systems to
which this approach is applicable, however, is limited due to the
number of terms in the reference state scaling exponentially with
the number of degrees of freedom.

We will now propose a simplification of Eq$.\ $(\ref{eq:HYB_sep})
that has a similar effect without needing a filter comprising such
a potentially high number of terms. First, we approximate the purely
TG parts of the integrand by their analytical harmonic oscillator
values
\begin{align}
\frac{1}{\left[\det\left(\mathbf{A}_{\text{HO}}\left(t\right)+\mathbf{A}_{\text{HO}}^{*}\left(t\right)\right)\right]^{1/4}} & \approx\left(2\hbar\right)^{F_{\text{tg}}/2}\label{eq:HYB_HO_detA}\\
\mathbf{b}_{t,\text{HO}}^{\text{T}}\left(\mathbf{A}_{\text{HO}}\left(t\right)+\mathbf{A}_{\text{HO}}^{*}\left(t\right)\right)^{-1}\mathbf{b}_{t,\text{HO}} & \approx0,\label{eq:HYB_HO_bTAb}
\end{align}
that are derived shortly in App.$\ $\ref{sec:mixed_quantities}.
This already results in a considerably simpler form for Eq$.\ $(\ref{eq:HYB_sep}),
\begin{align}
I(E) & =\frac{1}{\left(2\pi\hbar\right)^{F_{\text{hk}}}}\frac{1}{2\pi\hbar T}\int\text{d}\mathbf{p}_{\text{hk}}\left(0\right)\int\text{d}\mathbf{q}_{\text{hk}}\left(0\right)\nonumber \\
 & \times\left|\int_{0}^{T}\text{d}t\:\text{e}^{\text{i}\left[Et+\phi_{t}\left(\mathbf{p}\left(0\right),\mathbf{q}\left(0\right)\right)+S_{t}\left(\mathbf{p}\left(0\right),\mathbf{q}\left(0\right)\right)\right]/\hbar}\right.\nonumber \\
 & \times\left.\left\langle \mathbf{p}_{\text{eq}},\mathbf{q}_{\text{eq}}\middle|\mathbf{p}\left(t\right),\mathbf{q}\left(t\right)\right\rangle \right|^{2}.\label{eq:HYB_sep_inter}
\end{align}
In a second step, we choose the reference state $\left\langle \mathbf{p}_{\text{eq}},\mathbf{q}_{\text{eq}}\right|$
as a filter in the spirit of Eq$.\ $(\ref{eq:filter_MCTASCIVR}),
but we define it in a different, partially time-dependent fashion,
\begin{align}
\begin{split}\left\langle \mathbf{p}_{\text{eq}},\mathbf{q}_{\text{eq}}\right|\rightarrow & \left(\prod_{j=1}^{F_{\text{hk}}}\left\langle p_{\text{eq},\text{hk},j},q_{\text{eq},\text{hk},j}\right|\right)\left(\prod_{k=1}^{F_{\text{tg}}}\left\langle p_{\text{tg},k}(t),q_{\text{tg},k}(t)\right|\right)\\
 & =\left\langle \mathbf{p}_{\text{eq},\text{hk}},\mathbf{q}_{\text{eq},\text{\text{hk}}}\right|\left\langle \mathbf{p}_{\text{tg}}\left(t\right),\mathbf{q}_{\text{\text{tg}}}\left(t\right)\right|.
\end{split}
\end{align}
Since the time-dependent part is exactly the complex conjugate of
the thawed Gaussian contribution to the time-evolved wavepacket, it
cancels this part of the overlap in Eq.$\ $(\ref{eq:HYB_sep_inter}).
The final simplified approach to the mixed TA-SCIVR, which we will
refer to as SAM-TA-SCIVR or simply SAM, is thus 
\begin{align}
I(E)= & \ \frac{1}{\left(2\pi\hbar\right)^{F_{\text{hk}}}}\frac{1}{2\pi\hbar T}\int\text{d}\mathbf{p}_{\text{hk}}\left(0\right)\int\text{d}\mathbf{q}_{\text{hk}}\left(0\right)\nonumber \\
 & \times\left|\int_{0}^{T}\text{d}t\:\text{e}^{\text{i}\left[Et+\phi_{t}\left(\mathbf{p}\left(0\right),\mathbf{q}\left(0\right)\right)+S_{t}\left(\mathbf{p}\left(0\right),\mathbf{q}\left(0\right)\right)\right]/\hbar}\right.\nonumber \\
 & \times\left.\left\langle \mathbf{p}_{\text{eq},\text{\text{hk}}},\mathbf{q}_{\text{eq},\text{\text{hk}}}\middle|\mathbf{p}_{\text{\text{hk}}}\left(t\right),\mathbf{q}_{\text{hk}}\left(t\right)\right\rangle \right|^{2}.\label{eq:SAM_TA_SCIVR}
\end{align}
The remaining quantities in the integrand, namely, the action
\begin{align}
S_{t}\left(\mathbf{p}(0),\mathbf{q}(0)\right) & =S_{t}\left(\mathbf{p}_{\text{hk}}(0),\mathbf{q}_{\text{hk}}(0);\mathbf{p}_{\text{eq},\text{tg}},\mathbf{q}_{\text{eq},\text{tg}}\right)\label{eq:SAM_TA_action}
\end{align}
as well as the prefactor phase
\begin{align}
\phi_{t}\left(\mathbf{p}(0),\mathbf{q}(0)\right)= & \phi_{t}\left(\mathbf{p}_{\text{hk}}(0),\mathbf{q}_{\text{hk}}(0);\mathbf{p}_{\text{eq},\text{tg}},\mathbf{q}_{\text{eq},\text{tg}}\right),
\end{align}
are not affected by the simplifications. We stress that these quantities
already ``live'' in a reduced dimensionality: while their classical
evolution depends on the initial conditions of all DOFs, only the
HK initial coordinates are variables of the phase space sampling.
The TG DOFs' initial positions and momenta are fixed and can therefore
be seen as parameters of the phase space integration. In this way,
the integration as well as the integrand are restricted to the HK
part of phase space.

By comparison with the original TA-SCIVR \equationabb(\ref{eq:HK_sep}),
one can see that Eq.$\ $(\ref{eq:SAM_TA_SCIVR}) is indeed the original
time-averaged result with the sampling reduced to a selection of degrees
of freedom, while the remaining degrees of freedom are always taken
to be initially at the center of the reference state as in the mixed
approach. The classical dynamics is still the full dynamics of system
and environment combined.

The effect of this drastic simplification of the M-TA-SCIVR is investigated
analytically for two uncoupled harmonic oscillators in App.$\ $\ref{sec:HO-analytical},
and for two different numerical applications in the following sections.
As we will see, it does indeed serve as a filter by virtue of removing
odd bath peaks, in particular the first harmonics of the bath oscillators.
This results in a significant reduction of noise in the spectra, especially
when going to higher bath dimensionality. The weight and accuracy
of the HK peaks, on the other hand, is not affected. As a slight drawback,
even bath excitations are reflected at the system peaks and show up
as ghost peaks in the spectrum. Since we are not interested in bath
excitations anyway, and because these artifacts are always less prominent
than neighboring system excitations, we believe this additional inaccuracy
is a small price to pay, compared to the huge benefit of recovering
meaningful information from an otherwise unreadable spectrum.

\section{Results and Discussion\label{sec:results}}

\subsection{Morse oscillator coupled to harmonic oscillators}

Our first test system will be the Caldeira-Leggett Hamiltonian
\begin{align}
H= & \frac{p_{\text{s}}^{2}}{2m_{\text{s}}}+V_{\text{s}}(s)+\sum\limits _{i=1}^{F_{\text{b}}}\left[\frac{p_{i}^{2}}{2}+\frac{\omega_{i}^{2}}{2}\left(y_{i}+\frac{c_{i}}{\omega_{i}^{2}}\left(s-s_{\text{eq}}\right)\right)^{2}\right],
\end{align}
and we use a Morse potential with the parameters of molecular iodine
\cite{Buchholz_Ceotto_MixedSC_2016} as the system, 
\begin{align}
V_{\text{s}}(s)= & D_{\text{e}}\left(1-\text{e}^{-\alpha\left(s-s_{\text{eq}}\right)}\right)^{2}.
\end{align}
The bath is characterized by a discretized Ohmic spectral density,\cite{Goletz_Grossmann_Decoherence_2009,Buchholz_Ceotto_MixedSC_2016,Ceotto_Buchholz_MixedSC_2017,Goletz2010}
resulting in frequencies 
\begin{align}
\omega_{i} & =-\omega_{\text{c}}\ln\left(1-\frac{i(1-\text{e}^{-\omega_{\text{max}}/\omega_{\text{c}}})}{F_{\text{b}}}\right).
\end{align}
 We use a small cutoff and maximum frequency, $\omega_{\text{c}}=\omega_{\text{max}}=0.2\ \omega_{\text{e}}$,
where $\omega_{\text{e}}$ is the harmonic approximation frequency
of the Morse oscillator. The dimensionless effective coupling strength
is $\eta_{\text{eff}}=0.2$. This situation is similar in terms of
frequency difference to the experimentally investigated iodine molecule
in a krypton environment that we will discuss below. First, the environment
comprises four oscillators such that comparison to the other semiclassical
approaches is possible. Each degree of freedom is initially positioned
at its potential minimum with initial momentum corresponding to its
ground state energy, $p_{i}=\sqrt{m_{i}\omega_{i}},$ with $\omega_{1}=\omega_{\text{e}}$
for the system coordinate. $10^{4}$ trajectories are sufficient to
reach convergence with respect to peak positions. Peak amplitudes
may differ to a small degree with more trajectories added to the phase
space integration. However, amplitudes as well as peak shapes are
not our main interest, because already the original separable approximation
by Kaledin and Miller\cite{Kaledin_Miller_TAmolecules_2003,Kaledin_Miller_Timeaveraging_2003}
introduces significant quantitative inaccuracies for these quantities.

\begin{figure}
\centering{}\includegraphics[scale=0.9]{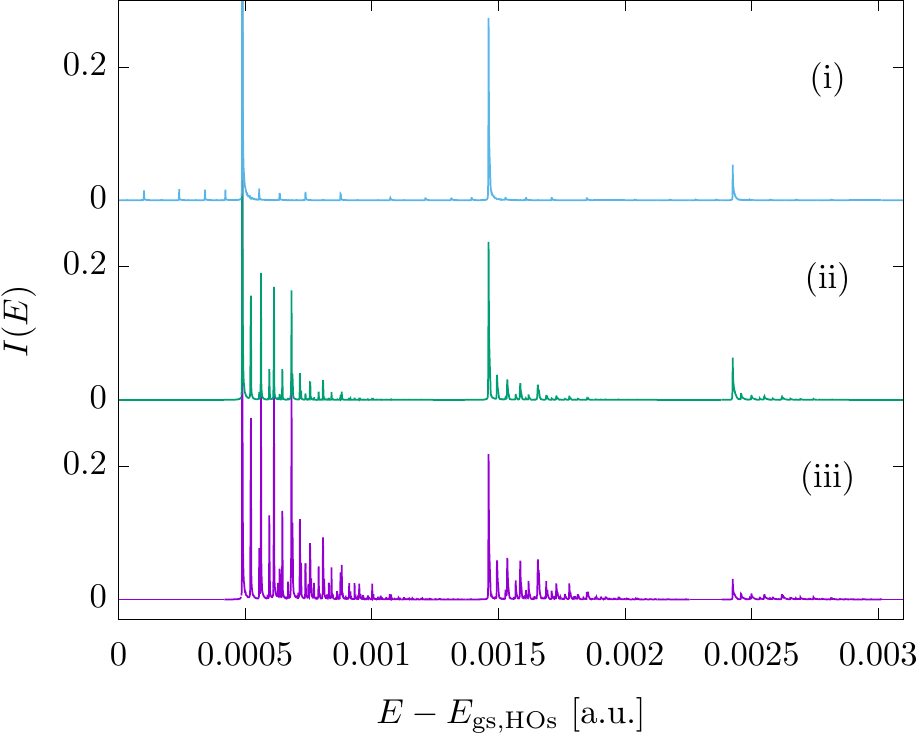}\caption{\label{fig:Morse_CL_weak}Morse oscillator coupled to four harmonic
oscillators with $\omega_{\text{c}}=\omega_{\text{max}}=0.2\ \omega_{\text{e}}$
and $\eta_{\text{eff}}=0.2$. From top to bottom: SAM-TA-SCIVR according
to \equationabb(\ref{eq:SAM_TA_SCIVR}) (blue line, (i)), M-TA-SCIVR
as in \equationabb(\ref{eq:HYB_sep}) (green line, (ii)) and full
TA-SCIVR as in \equationabb(\ref{eq:HK_sep}) (violet line, (iii)).
All elastic peaks are normalized to one, and the spectra overlap to
make higher excitations more visible.}
\end{figure}

Results are shown in \figureabb\ref{fig:Morse_CL_weak}, where the
ground state energies of the bath HOs, $E_{\text{gs},\text{HO}}=\sum_{i}\omega_{i}/2$,
have been subtracted. The degree of approximation decreases from top
to bottom, with SAM-TA-SCIVR according to \equationabb(\ref{eq:SAM_TA_SCIVR})
shown in blue (i), M-TA-SCIVR as in \equationabb(\ref{eq:HYB_sep})
in green (ii) and full TA-SCIVR as in \equationabb(\ref{eq:HK_sep})
with violet lines (iii). All three approaches agree within frequency
resolution in terms of peak positions. As expected and as desired,
bath excitations are very much suppressed by the SAM-TA-SCIVR method.
Unlike in the two reference spectra, some small ghost peaks appear
in the SAM result, for example to the left of (and therefore at unphysical
smaller energy than) the elastic peak. As shown analytically in App.$\ $\ref{sec:HO-analytical}
for two uncoupled harmonic oscillators, these ghost peaks are second
excitations of the bath modes reflected at the elastic peak. Upon
close inspection, the same behavior can be observed for all higher
excitations of the system. The ghost peaks are not a problem for the
interpretation of the spectrum, as they are far smaller than the respective
HK peak they are close to. In addition, they can be identified from
their position, which is always an integer multiple of a bath frequency
(or a combination thereof) to the left of a system excitation if the
bath modes are sufficiently harmonic.

\begin{figure}
\begin{centering}
\includegraphics{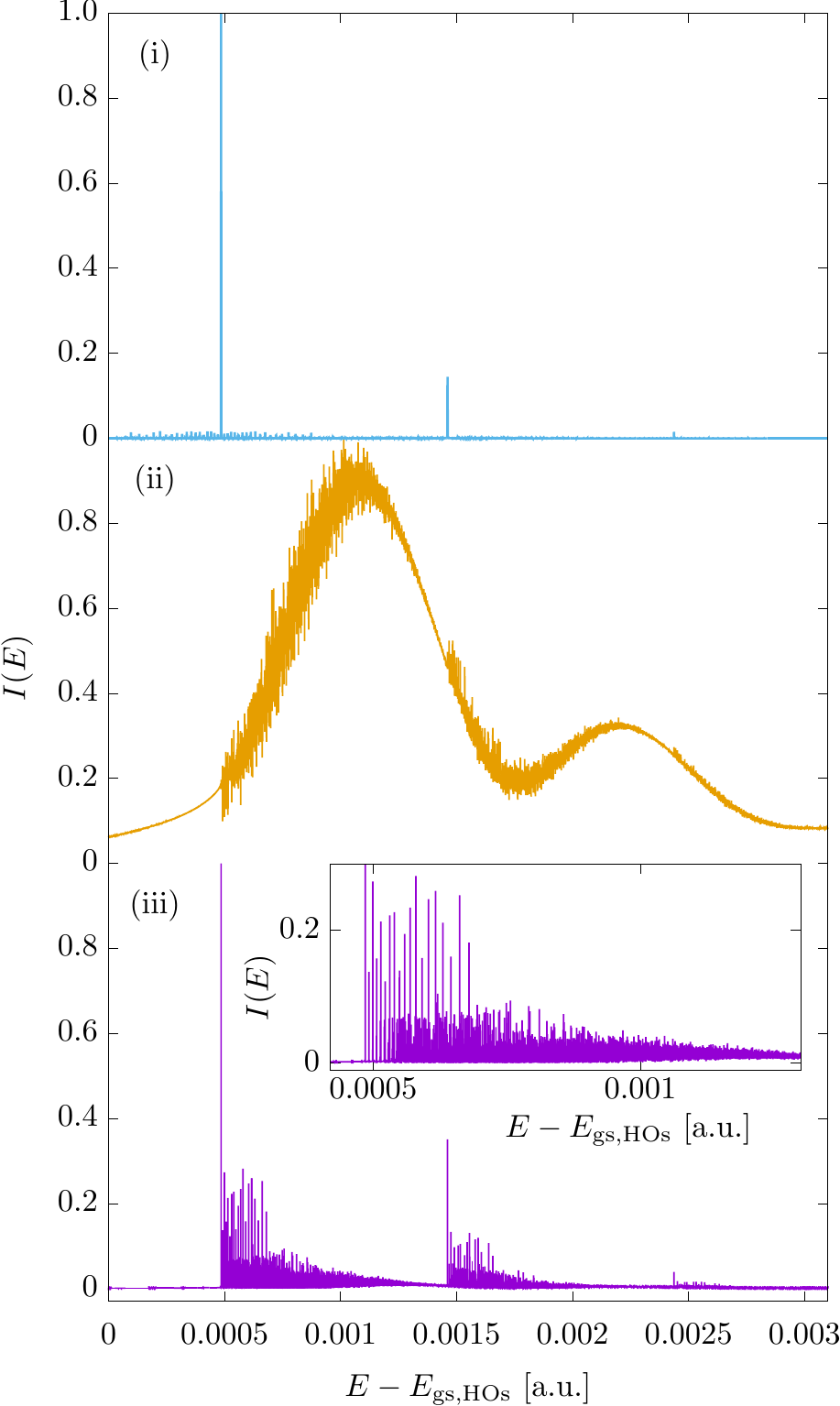}
\par\end{centering}
\caption{\label{fig:Morse-CL-big}Morse oscillator coupled to 18 harmonic oscillators
with $\omega_{\text{c}}=\omega_{\text{max}}=0.2\ \omega_{\text{e}}$
and $\eta_{\text{eff}}=0.2$. From top to bottom: SAM-TA-SCIVR according
to \equationabb(\ref{eq:SAM_TA_SCIVR}) (blue line, (i)) with $2^{15}$
propagation time steps, full TA-SCIVR as in \equationabb(\ref{eq:HK_sep})
with $2^{15}$ propagation time steps (orange line, (ii)), and full
TA-SCIVR with $2^{20}$ propagation time steps (violet line, (iii)).
The respective highest peaks are normalized to one.}
\end{figure}

While the spectrum with five weakly coupled, off-resonant oscillators
already contains a lot of bath excitations, all of these peaks can
be assigned without difficulty. In the next example, we show a situation
where this is not the case any more. The bath has still the same parameters,
but now comprises 18 instead of four harmonic oscillators. We restrict
the calculation to a single trajectory, which is sufficient to demonstrate
the main challenge arising from this higher number of DOFs. The initial
conditions are the same as before, with all DOFs centered at $(p_{\text{eq},i},q_{\text{eq},i})$.
Results are shown in Fig.$\ $\ref{fig:Morse-CL-big}, with the SAM-TA-SCIVR
result (blue line, (i)) on top, and the two reference HK calculations
with different propagation times are below ((ii) and (iii)). For this
higher number of bath DOFs, we see that the propagation time from
the previous 5D example with $2^{15}$ steps, which leads to a numerical
energy resolution $\Delta E=6\times10^{-7}\ \text{a.u.}\ \left(0.13\ \text{cm}^{-1}\right)$,
is not sufficient any more to obtain a well-resolved spectrum. This
is illustrated by the reference HK calculation with this number of
time steps (Fig.$\ $\ref{fig:Morse-CL-big}(ii), orange line), where
the much higher number of bath excitations leads to a quasi-continuous
spectrum that is much broader than before and does not allow for an
unambiguous attribution of peaks. It is possible to recover a discrete
spectrum by significantly increasing the length of the propagation
and thereby the energy resolution. Results of the same reference TA-SCIVR
calculation with $2^{20}$ instead of $2^{15}$ time steps are reported
in the bottom spectrum of Fig.$\ $\ref{fig:Morse-CL-big} (violet
line, (iii)). Here, the system excitations can be seen clearly, and
the bath peaks are very dense but discrete. As we go from high energy
resolution in Fig.$\ $\ref{fig:Morse-CL-big}(iii) to the lower energy
resolution in Fig.$\ $\ref{fig:Morse-CL-big}(ii), distinct contributions
from the higher resolution can now coincide in the same energy bin.
Since the density of bath peaks gets higher far away from the system
excitation, as illustrated by the inset in Fig.$\ $\ref{fig:Morse-CL-big},
the lower resolution introduces an artificial bias that overestimates
relative peak weights in these regions of high bath peak density.
Conversely, the system excitations are underestimated and likely to
be absorbed in the quasi-continuum. Given that the phase space integrands
in Eqs. (\ref{eq:HK_sep}) and (\ref{eq:HYB_sep}) are positive definite,
the phase space average does not resolve this issue. Simply prolonging
the propagation time, on the other hand, recreates a discrete spectrum,
but this is by no means a feasible general solution, as much longer
propagation times are usually prohibitively expensive and may increase
the likelihood of numerical instability.

Instead, the inherent filter of SAM-TA-SCIVR (blue line, (i) in Fig.$\ $\ref{fig:Morse-CL-big})
offers a numerically cheap solution. With the same lower number of
$2^{15}$ time steps as in panel (ii), we obtain a completely different
picture. By removing contributions from first-order bath excitations,
the old hierarchy of prominent system peaks and very small bath excitations
from Fig.$\ $\ref{fig:Morse_CL_weak} is restored. Compared to the
higher accuracy calculation in Fig.$\ $\ref{fig:Morse-CL-big}(iii),
it is evident that the location of the system energies is reproduced
exactly. Given the higher number of bath oscillators, the number of
ghost peaks is getting bigger as well. However, at least in this weakly
coupled example, they are again the same order of magnitude as their
accurate counterparts and therefore easily distinguished from the
system excitations.

\subsection{Molecular iodine embedded in krypton}

Having introduced SAM-TA-SCIVR as a useful tool for the analysis of
high-dimensional spectra, we now turn to an experimentally investigated
system, namely, iodine in a krypton environment. Iodine surrounded
by noble gas atoms has been used as a test system for a number of
semiclassical approaches, for example the study of vibrational quantum
coherence of iodine in argon clusters using a forward-backward IVR.\cite{Tao_Miller_I2Ar_2009_1,Tao_Miller_I2Ar_2009_2,Tao_Feng_2013}
Another study of the loss of coherence of iodine in a krypton environment
has already established the hybrid formalism as an appropriate tool
for the investigation of this system.\cite{Buchholz_Jungwirth_CondensedPhase_2012}
Here, we are interested in the change of the iodine vibrational spectrum
by the surrounding krypton atoms. Experimentally, it has been found
that the iodine spectrum undergoes a redshift, from gas phase\cite{Bihary_Apkarian_VSCF_2001}
harmonic frequency $\omega_{\text{e}}=214.6$ $\text{cm}{}^{-1}$
and anharmonicity $\omega_{\text{e}}x_{\text{e}}=$0.627 $\text{cm}{}^{-1}$
to $\omega_{\text{e}}=211.6$ $\text{cm}{}^{-1}$ and $\omega_{\text{e}}x_{\text{e}}=$0.658
$\text{cm}{}^{-1}$ when embedded in krypton \cite{Karavitis2005},
see \tableabb\ref{tab:ikr_num_results}.

\subsubsection{Model: Dynamic Cell with Rigid Walls }

\begin{figure}
\begin{centering}
\includegraphics[scale=0.4]{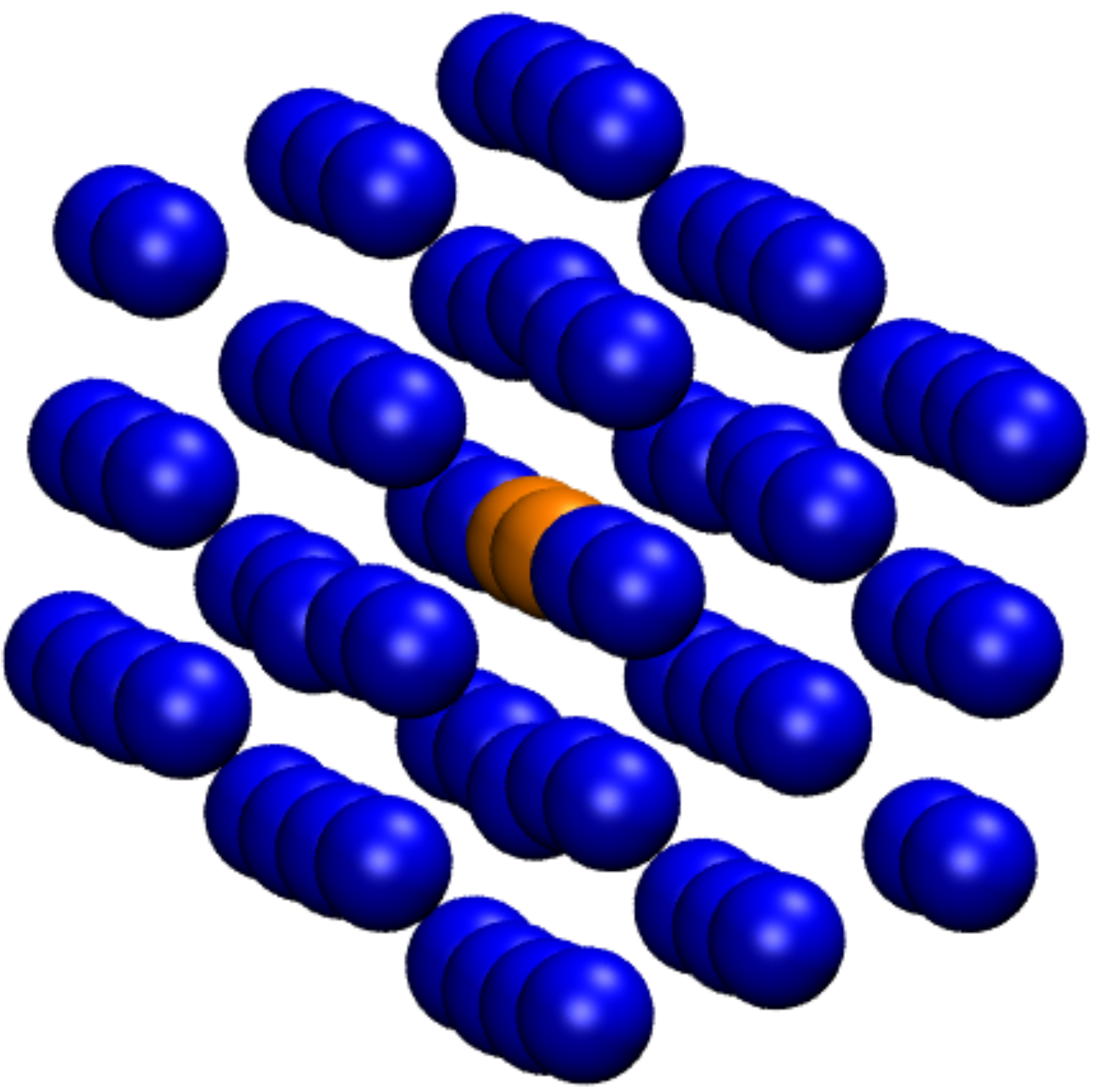}
\par\end{centering}
\caption{The iodine molecule (orange) in two flexible layers of krypton atoms
(blue) after geometry optimization. Not shown is the fixed outermost
layer of krypton atoms.\label{fig:I2Kr27}}
\end{figure}

As shown in a closely related investigation of iodine in an argon
matrix\cite{Bihary_Apkarian_VSCF_2001}, there are two important caveats
when it comes to spectral calculations of iodine in a rare gas environment.
First, one has to choose a suitable matrix environment to reproduce
the rare gas geometry faithfully, using a sufficient number of layers
around the host molecule. For iodine in argon, four such layers were
necessary for convergence with respect to the iodine frequency shift,
corresponding to 448 argon atoms. Of these, however, only the two
inner shells were taken to be mobile, while the two outer layers were
fixed during the propagation; this choice of boundary conditions is
called Dynamical Cell with Rigid Walls by the authors of Ref. \onlinecite{Bihary_Apkarian_VSCF_2001}.
We will use the same approach, but restrict the environment to just
three layers with 218 atoms for the classical geometry optimization,
where the outermost is fixed and the two inner ones, comprising 72
krypton atoms, are mobile. The iodine molecule is placed inside the
face-centered cubic (fcc) krypton lattice by replacing two nearest-neighbor
atoms. Then, we perform a geometry optimization for the iodine as
well as the mobile krypton atoms, while the outer, fixed krypton atoms
serve as containment. The minimum energy geometry is presented in
\figureabb \ref{fig:I2Kr27}, where iodine atoms are orange and the
flexible krypton atoms are blue. As a result, only a few atoms from
the innermost shell are notably shifted. The subsequent normal mode
analysis is performed only for the 74 flexible atoms depicted in \figureabb
\ref{fig:I2Kr27} with the TrajLab software. \cite{Schmidt2014}

As a second important point, it has been demonstrated that the halogen-rare
gas interaction potential is essential for getting the accurate iodine
bond softening which leads to the redshift. While an anisotropic interaction
of the form
\begin{align}
V_{ik}(\mathbf{R}_{ik},\mathbf{R}_{12}) & =\left(\cos\theta_{ik}\right)^{2}V_{\text{\ensuremath{\Sigma},}ik}\left(R_{ik}\right)+\left(\sin\theta_{ik}\right)^{2}V_{\text{\ensuremath{\Pi},}ik}\left(R_{ik}\right)\label{eq:I-Kr-anisotropic}
\end{align}
yields an even quantitatively accurate frequency shift for iodine
in argon, \cite{Bihary_Apkarian_VSCF_2001} other (simpler) analytic
interactions result in no shift at all or even a blueshift of the
iodine frequencies. In the above equation, the index $i$ designates
one of the two iodine atoms, while index $k$ stands for a krypton
atom. The angle $\theta_{ik}$ is the angle between $\mathbf{R}_{ik}$
and the iodine-iodine vector $\mathbf{R}_{12}$. The total potential
is modeled as a sum over two-particle interactions
\begin{align}
V(\mathbf{R}_{1},\dots,\mathbf{R}_{N})= & \sum_{i<j}V_{ij},
\end{align}
where we use \equationabb(\ref{eq:I-Kr-anisotropic}) for the iodine-krypton
interaction with Morse potentials $V_{\text{\ensuremath{\Sigma}}}$
and $V_{\Pi},$ a simple Morse potential for the iodine intramolecular
interaction and a Lennard-Jones (LJ) potential for the krypton-krypton
interaction. All potential parameters are collected in \tableabb\ref{tab:I2KrN-parameters}. 

\begin{table}
\begin{tabular}{cccc}
\hline 
\noalign{\vskip\doublerulesep}
Morse interaction & $D_{\text{e}}\left[\text{cm}^{-1}\right]$ & $s_{\text{e}}\left[\text{Å}\right]$ & $\alpha\left[\text{Å}^{-1}\right]$\tabularnewline[\doublerulesep]
\hline 
I-I\cite{Bihary_Apkarian_VSCF_2001} & 18357 & 2.666 & 1.536\tabularnewline
I-Kr ($\Sigma$)\cite{Ovchinnikov_Apkarian_Condensedphase_1996} & 287 & 3.733 & 1.49\tabularnewline
I-Kr ($\Pi$)\cite{Ovchinnikov_Apkarian_Condensedphase_1996} & 126 & 4.30 & 1.540\tabularnewline
\hline 
\noalign{\vskip\doublerulesep}
LJ interaction & $\epsilon\left[\text{cm}^{-1}\right]$ & $\sigma\left[\text{Å}\right]$ & \tabularnewline[\doublerulesep]
\hline 
Kr-Kr\cite{Zadoya1994,Ovchinnikov_Apkarian_Condensedphase_1996} & 138.7 & 3.58 & \tabularnewline
\hline 
\end{tabular}\caption{Parameters for the different two-particle interactions of the iodine-krypton
potential. \label{tab:I2KrN-parameters}}
\end{table}

\subsubsection{Reproducing the condensed phase quantum effects by adding degrees
of freedom}

After geometry optimization and subsequent normal mode analysis, we
find that, similar to iodine in argon, the resulting shifted harmonic
frequency of the iodine vibration is already very close to the experimental
result, $211.8\ \text{cm}^{-1}$ against $211.6\ \text{cm}^{-1}$.
The major portion of the shift from the gas phase result $214.6\ \text{cm}^{-1}$
is thus a classical effect due to the rearrangement of iodine molecule
and krypton atoms, and in particular the stretching of the iodine
bond, during the geometry optimization. 

In order to find the remaining contribution to the redshift due to
the quantum dynamical interaction of the iodine molecule with its
krypton environment and to include overtones, we perform SAM calculations
with different numbers of normal coordinates included into the dynamics.
We have to use $2^{17}$ semiclassical time steps with 2 classical
substeps of length $\left(2\pi/\omega_{\text{e}}\right)/120$, corresponding
to a frequency grid spacing of $4.4\times10^{-7}\ \text{a.u.}$ ($0.1\ \text{cm}^{-1}$),
in order to faithfully resolve possible differences. The standard
phase space sampling even of a single HK DOF becoming unfeasible as
the number of normal modes approaches three digits because the SAM
expression (\ref{eq:SAM_TA_SCIVR}) still requires the propagation
of the full HK prefactor. Thus, the main computational challenge is
the calculation of the second derivatives in normal coordinates, which
brings about a computational time of a few days for one trajectory
with about 100 DOFs. As a consequence, we resort to the MC-SCIVR idea
that each single classical trajectory reproduces the part of the spectrum
exactly that is closest to its own energy, and run only six trajectories
with different initial momenta for the iodine vibrational coordinate.
Each of these initial momenta corresponds to the energy of one of
the first six excited states of iodine, which we can get approximately
from a multiple-trajectory calculation of iodine in the rigid krypton
cage. 

\begin{figure}
\begin{centering}
\includegraphics{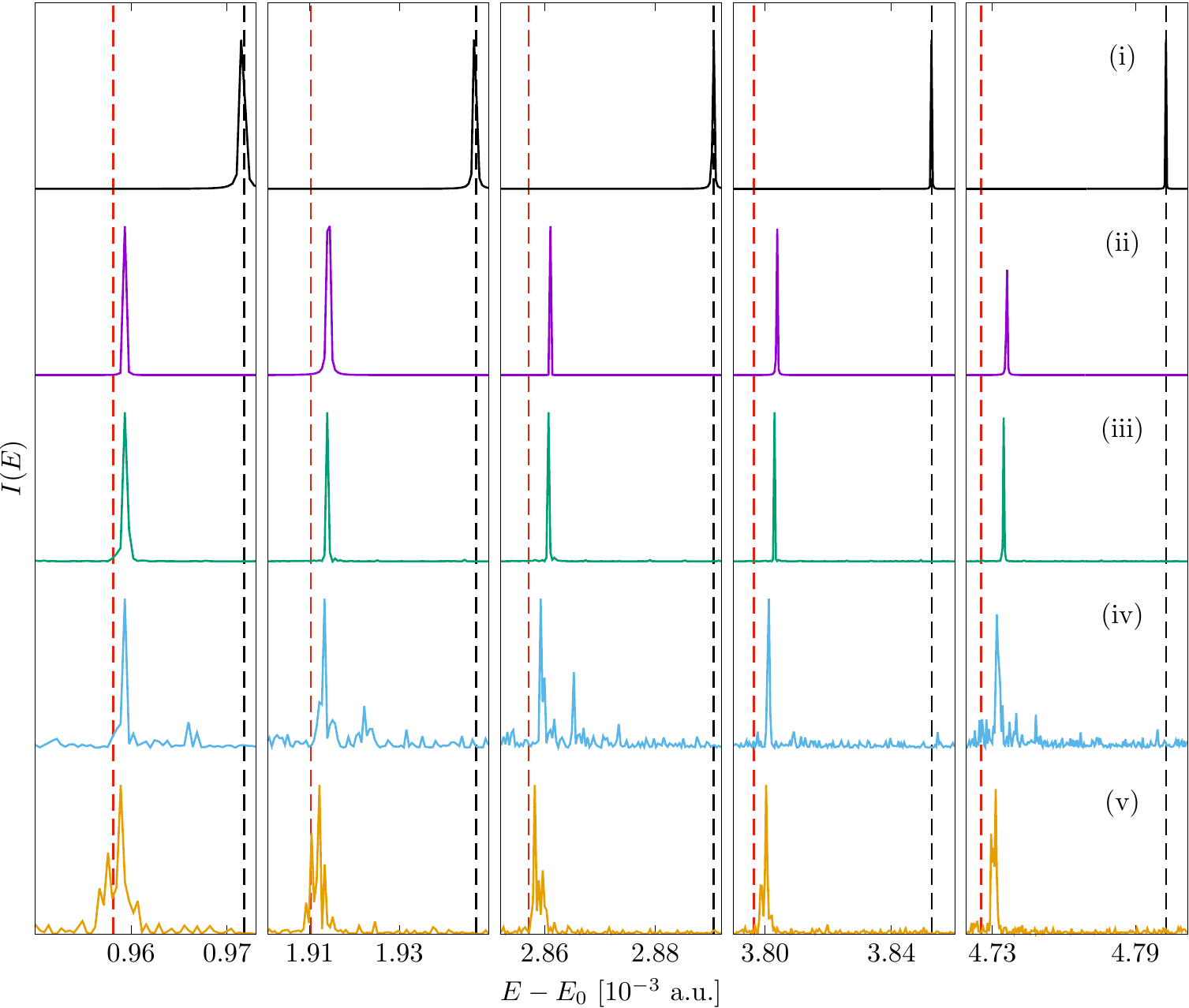}
\par\end{centering}
\caption{SAM-TA-SCIVR spectra for iodine in krypton for different numbers of
normal modes included in the dynamical calculation. From left to right,
the regions around the first through the fifth excited vibrational
state of iodine are shown. The energy of the respective elastic peak
is subtracted on the abscissa to make calculations comparable. From
top to bottom: gas phase result (solid black lines, (i)), iodine vibration
only in a rigid krypton cage (violet, (ii)), and spectra from calculations
with 32 (green, (iii)), 60 (blue, (iv)), and 108 (orange, (v)) flexible
normal modes. The dashed lines show the analytic positions of the
experimental results for gas-phase iodine\cite{Bihary_Apkarian_VSCF_2001}
(black) and redshifted iodine in krypton \cite{Karavitis2005} (red).
\label{fig:ikr_SAM_spectra}}
\end{figure}

\begin{table}
\begin{tabular}{ccc}
\hline 
\noalign{\vskip\doublerulesep}
Experiment & $\omega_{\text{e}}\left[\text{cm}^{-1}\right]$ & $\omega_{\text{e}}x_{\text{e}}\left[\text{cm}^{-1}\right]$\tabularnewline[\doublerulesep]
\hline 
gas phase $\text{I}_{2}$\cite{Bihary_Apkarian_VSCF_2001} & 214.6 & 0.627\tabularnewline
$\text{I}_{2}$ in Kr\cite{Karavitis2004} & 211.3 & 0.652\tabularnewline
$\text{I}_{2}$ in Kr\cite{Karavitis2005} & 211.6 & 0.658\tabularnewline
\hline 
\hline 
\noalign{\vskip\doublerulesep}
Numerical results & $\omega_{\text{e}}\left[\text{cm}^{-1}\right]$ & $\omega_{\text{e}}x_{\text{e}}\left[\text{cm}^{-1}\right]$\tabularnewline[\doublerulesep]
\hline 
normal mode analysis & 211.8 & \textendash{}\tabularnewline
$\text{I}_{2}$ vibration in rigid Kr cage & 211.8 & 0.62\tabularnewline
32 DOFs & 211.8 & 0.63\tabularnewline
60 DOFs & 211.7 & 0.64\tabularnewline
108 DOFs & 211.6 & 0.63\tabularnewline
\hline 
\end{tabular}

\caption{Spectroscopic parameters of the molecular iodine Morse potential.
All numerical results have been obtained from Birge-Sponer fits to
the first six iodine eigenenergies from the SAM-TA-SCIVR spectrum.
The number of DOFs indicates how many of the 216 internal DOFs of
the two-shell iodine-krypton cluster have been considered in the dynamics.\label{tab:ikr_num_results}}
\end{table}

In our SAM calculations, which are summarized in \figureabb \ref{fig:ikr_SAM_spectra}
and \tableabb\ref{tab:ikr_num_results}, we always treat the iodine
vibration as the only HK DOF. The first calculation (solid violet
line, (ii), in \figureabb \ref{fig:ikr_SAM_spectra}) comprises only
this one DOF, i.e. all krypton atoms are rigid. All excited peaks
from the five single-trajectory calculations are very clean, as there
are only iodine excitations in the respective spectra. A Birge-Sponer
fit reveals that the harmonic frequency is indeed identical to the
eigenvalue of the normal mode, as shown in \tableabb\ref{tab:ikr_num_results}.
The anharmonicity $x_{\text{e}}\omega_{\text{e}}$ is clearly closer
to the gas phase result than to the redshifted result in krypton.
Increasing the dimensionality of the system, we now take into account
all 32 fully symmetric normal modes (green lines, (iii), in \figureabb
\ref{fig:ikr_SAM_spectra}). All bath DOFs carry initial momentum
corresponding to the harmonic frequency of the respective mode. We
thus need to employ the SAM filter, which works as intended in fully
removing all bath excitations at least in close proximity to the system
peaks. These peaks are already shifted a little bit towards the redshifted
experimental result, which is also reflected in the results of the
Birge-Sponer fit. By adding more normal modes to the dynamical calculation,
we can see that this trend continues. With 59 TG normal modes with
the same initial excitation as before (blue lines, (iv)), we still
find system peaks that can clearly be identified as such, but also
a certain amount of noise that is no longer filtered out completely.
The growing number of normal modes influencing the dynamics of the
iodine vibration causes another quite significant shift towards the
experimental result. 

Going up to include 108 normal modes, we finally approach the limits
of SAM. We have given only 60 modes initial excitation corresponding
to the respective mode's ground state energy $\omega_{i}$ while the
remaining modes have zero initial momentum. Higher initial energy
of the remaining bath DOFs would bring about significant excitation
of some higher order bath states, such that the corresponding peaks
are not filtered sufficiently by the SAM approach any more. In spite
of this limitation, we still see an improvement of the excited peak
positions in the case of 108 normal modes (orange lines, (v), in \figureabb
\ref{fig:ikr_SAM_spectra}). This also shows in the fitted value for
$\omega_{\text{e}}$, which is closest to the experimental results
(\tableabb\ref{tab:ikr_num_results}).

\section{Conclusions and Outlook\label{sec:conc}}

In this work, we have presented a simplified version of the mixed
time-averaging SCIVR. The underlying idea of M-TA-SCIVR is a hybrid
description of phase space, where a small part is described on the
HK level of accuracy, while the remaining environment is treated with
TG. The SAM-TA-SCIVR builds upon this idea, but takes another approximative
step by using the exact harmonic oscillator results for the TG part
of the phase space integrand. This leads to a considerably simplified
expression, reminiscent of the original HK TA-SCIVR with a reduced
phase space sampling. The motivation for this approximation was to
find an approach that is still accurate for the HK DOFs, but at the
same time removes bath excitations from the spectrum.

In the application to a Morse oscillator coupled to a Caldeira-Leggett
bath, we have seen for a small number of bath degrees of freedom that
SAM indeed reproduces the HK peaks on the same level of accuracy as
reference HK or mixed calculations. The decisive advantage over the
formally more accurate approaches is the removal of odd harmonics
of the bath oscillators. For twenty bath oscillators, this filter
effect proved its value by recreating a clear spectrum from the same
classical dynamics that yields excessively noisy spectra with the
reference methods. The only unfavorable characteristic of the SAM
spectra are some ghost peaks from the environmental DOFs. Since these
ghost peaks are both very small and appear at well-understood positions,
we do not consider them a problem for the applicability of the method.

Investigating an experimentally studied problem, namely, the redshift
of the iodine molecule embedded in a krypton environment, we first
saw that the shift of the harmonic approximation frequency $\omega_{\text{e}}$
of the iodine Morse potential is mainly an effect resulting from the
rearrangement of iodine and krypton atoms in a classical geometry
optimization. We then used SAM to find the effect of the krypton environment
on the excited iodine vibrational energies and to see the change in
iodine energies due to the dynamical interaction with the environment.
The SAM approach allowed to include up to 108 vibrational DOFs in
the calculation, where either all or at least the majority of the
bath modes carry initial excitation. In spite of all interactions
in this system being anharmonic, the bath filter still works as intended
and we can easily identify the system excitations. In the model with
two flexible inner krypton shells, contained by a fixed outer layer,
we could show that adding more and more normal modes to the calculation
systematically improves the result towards very good quantitative
agreement with experimental findings. Thus, the increasingly complex
environment appropriately captures the effect of the system-bath dynamics
on the iodine spectrum, both for the fundamental frequency and the
overtones. 

In the future, the SAM approach will be combined with Divide-and-Conquer
SCIVR\cite{ceotto_conte_DCSCIVR_2017,DiLiberto_Ceotto_Jacobiano_2018}
for tackling condensed phase systems using ab initio potentials. This
combination of methods will allow the system to be significantly higher
in dimensionality since the SAM approach properly embeds the system
into a condensed phase environment.

\section{Acknowledgements}

Michele Ceotto and Max Buchholz acknowledge financial support from
the European Research Council (ERC) under the European Union\textquoteright s
Horizon 2020 research and innovation programme (Grant Agreement No.
{[}647107{]} \textemdash{} SEMICOMPLEX \textemdash{} ERC-2014-CoG).
M.C. acknowledges also the CINECA and the Regione Lombardia award
under the LISA initiative (grant GREENTI) for the availability of
high performance computing resources. M.B. would like to thank his
son for providing the acronym for the method proposed in this paper.

\appendix

\section{Quantities From the Separable Mixed Expression\label{sec:mixed_quantities}}

For this paper to be self-contained, we briefly collect the terms
from the separable mixed approximation in Eq.$\ $(\ref{eq:HYB_sep})
that have not been defined in the main text. The $2F_{\text{tg}}\times2F_{\text{tg}}$
matrix $\mathbf{A}(t)$, which collects coefficients of the quadratic
deviations from the TG initial conditions, consists of four submatrices
defined as\cite{Ceotto_Buchholz_MixedSC_2017} 
\begin{align}
\begin{split}\mathbf{A}_{11}(t)= & \frac{1}{4}\mathbf{m}_{21}^{\text{T}}\left(t\right)\mathbf{\boldsymbol{\gamma}}\mathbf{m}_{21}\left(t\right)+\frac{1}{4\hbar^{2}}\mathbf{m}_{11}^{\text{T}}\left(t\right)\mathbf{\boldsymbol{\gamma}}^{-1}\mathbf{m}_{11}\left(t\right)\\
\mathbf{A}_{12}(t)= & \frac{1}{4}\mathbf{m}_{21}^{\text{T}}\left(t\right)\boldsymbol{\gamma}\mathbf{m}_{22}\left(t\right)+\frac{1}{4\hbar^{2}}\mathbf{m}_{11}^{\text{T}}\left(t\right)\boldsymbol{\gamma}^{-1}\mathbf{m}_{12}\left(t\right)\\
\mathbf{A}_{21}(t)= & \frac{1}{4}\mathbf{m}_{22}^{\text{T}}\left(t\right)\boldsymbol{\gamma}\mathbf{m}_{21}\left(t\right)+\frac{1}{4\hbar^{2}}\mathbf{m}_{12}^{\text{T}}\left(t\right)\boldsymbol{\gamma}^{-1}\mathbf{m}_{11}\left(t\right)\\
\mathbf{A}_{22}(t)= & \frac{1}{4}\mathbf{m}_{22}^{\text{T}}\left(t\right)\boldsymbol{\gamma}\mathbf{m}_{22}\left(t\right)+\frac{1}{4\hbar^{2}}\mathbf{m}_{12}^{\text{T}}\left(t\right)\mathbf{\boldsymbol{\gamma}}^{-1}\mathbf{m}_{12}\left(t\right).
\end{split}
\label{eq:HYB_A_matrix}
\end{align}
Prefactors of terms linear in the deviations are summarized in the
$2F_{\text{tg}}$-dimensional vector $\mathbf{b}(t)\equiv\left(\mathbf{b}_{1,t}^{\text{T}},\mathbf{b}_{2,t}^{\text{T}}\right)^{\text{T}}$
with subvectors 
\begin{align}
\begin{split}\mathbf{b}_{1,t}^{\text{T}}= & -\frac{1}{2}\left(\mathbf{q}\left(t\right)-\mathbf{q}\left(0\right)\right)^{\text{T}}\left[\mathbf{\boldsymbol{\gamma}}\mathbf{m}_{21}\left(t\right)+\frac{\text{i}}{\hbar}\mathbf{m}_{11}\left(t\right)\right]\\
 & -\frac{1}{2\hbar^{2}}\left(\mathbf{p}\left(t\right)-\mathbf{p}\left(0\right)\right)^{\text{T}}\left[\mathbf{\boldsymbol{\gamma}}^{-1}\mathbf{m}_{11}\left(t\right)-\text{i}\hbar\mathbf{m}_{21}\left(t\right)\right]\\
\mathbf{b}_{2,t}^{\text{T}}= & -\frac{1}{2}\left(\mathbf{q}\left(t\right)-\mathbf{q}\left(0\right)\right)^{\text{T}}\left[\mathbf{\boldsymbol{\gamma}}\mathbf{m}_{22}\left(t\right)+\frac{\text{i}}{\hbar}\mathbf{m}_{12}\left(t\right)\right]\\
 & -\frac{1}{2\hbar^{2}}\left(\mathbf{p}\left(t\right)-\mathbf{p}\left(0\right)\right)^{\text{T}}\left[\mathbf{\boldsymbol{\gamma}}^{-1}\mathbf{m}_{12}\left(t\right)-\text{i}\hbar\mathbf{m}_{22}\left(t\right)\right],
\end{split}
\label{eq:HYB_b_vector}
\end{align}
where we remind the reader that $\left(\mathbf{p}(t),\mathbf{q}(t)\right)$
is the trajectory starting at the mixed initial conditions $\left(\mathbf{p}(0),\mathbf{q}(0)\right)$
from Eq.$\ $(\ref{eq:HYB_initial}). The $\mathbf{m}_{ij}$ in the
two above equations are non-square $F\times F_{\text{tg}}$ submatrices
of the stability matrix, 
\begin{align}
\begin{split}\mathbf{m}_{11}(t)= & \frac{\partial\mathbf{p}(t)}{\partial\mathbf{p}_{\text{tg}}(0)},\quad\mathbf{m}_{12}(t)=\frac{\partial\mathbf{p}(t)}{\partial\mathbf{q}_{\text{\text{tg}}}(0)},\\
\mathbf{m}_{21}(t)= & \frac{\partial\mathbf{q}(t)}{\partial\mathbf{p}_{\text{tg}}(0)},\quad\mathbf{m}_{22}(t)=\frac{\partial\mathbf{q}(t)}{\partial\mathbf{q}_{\text{\text{tg}}}(0)}.
\end{split}
\label{eq:HYB_mij}
\end{align}
Assuming, for simplicity, a system of two uncoupled harmonic oscillators
with equations of motion 
\begin{align}
\begin{split}p_{i}(t) & =p_{\text{eq},i}\cos\omega_{i}t-m_{i}\omega_{i}q_{\text{eq},i}\sin\omega_{i}t\\
q_{i}(t) & =q_{\text{eq},i}\cos\omega_{i}t+\frac{p_{\text{eq},i}}{m_{i}\omega_{i}}\sin\omega_{i}t,
\end{split}
\label{eq:2HOs_eom}
\end{align}
and treating site 1 with HK and site 2 on the TG level, we get zero
for the respective first component of the stability submatrices, $m_{ij,1}=0$,
and the second components amount to
\begin{align}
\begin{split}m_{11,2}(t)= & \cos\omega_{2}t,\quad m_{12,2}(t)=-m_{2}\omega_{2}\sin\omega_{2}t,\\
m_{21,2}(t)= & \sin\omega_{2}t/(m_{2}\omega_{2}),\quad m_{22,2}(t)=\cos\omega_{2}t.
\end{split}
\label{eq:2HOs_mij}
\end{align}
With the usual $\gamma_{i}=m_{i}\omega_{i}/\hbar$, the matrix $\mathbf{A}$
becomes time independent,
\begin{align}
 & \mathbf{A}=\frac{1}{4}\left(\begin{array}{cc}
1/(\hbar m_{2}\omega_{2}) & 0\\
0 & m_{2}\omega_{2}/\hbar
\end{array}\right).
\end{align}
We thus find for this single uncoupled TG DOF
\begin{align}
\frac{1}{\left[\det\left(\mathbf{A}\left(t\right)+\mathbf{A}^{*}\left(t\right)\right)\right]^{1/4}} & =\left(2\hbar\right)^{1/2},
\end{align}
and this result can be easily generalized to the case of $F_{\text{tg}}$
uncoupled HOs to yield \equationabb(\ref{eq:HYB_HO_detA}). With
the inverse 
\begin{align}
\left(\mathbf{A}\left(t\right)+\mathbf{A}^{*}\left(t\right)\right)^{-1} & =\left(\begin{array}{cc}
2\hbar m_{2}\omega_{2} & 0\\
0 & 2\hbar/\left(m_{2}\omega_{2}\right)
\end{array}\right)
\end{align}
and the relation 
\begin{align}
b_{2,t} & =-\text{i}m_{2}\omega_{2}b_{1,t},
\end{align}
we obtain the one-dimensional case of \equationabb(\ref{eq:HYB_HO_bTAb}),
\begin{align}
\mathbf{b}_{t}^{\text{T}}\left(\mathbf{A}\left(t\right)+\mathbf{A}^{*}\left(t\right)\right)^{-1}\mathbf{b}_{t} & =2\hbar m_{2}\omega_{2}b_{1,t}^{2}+\frac{2\hbar}{m_{2}\omega_{2}}\left(\text{i}m_{2}\omega_{2}\right)^{2}b_{1,t}^{2}=0.
\end{align}
All of these results will be used in the application of SAM-TA-SCIVR
to the calculation of the spectrum of two uncoupled HOs in \appendixabb\ref{sec:HO-analytical}.

\section{Analytical Application of SAM-TA-SCIVR to Two Uncoupled Harmonic
Oscillators\label{sec:HO-analytical}}

The dynamics of two uncoupled HOs of unit mass with frequencies $\omega_{1}$
and $\omega_{2}$, respectively, is described by the Hamiltonian (in
atomic units)
\begin{align}
H & =\frac{p_{1}^{2}}{2}+\frac{p_{2}^{2}}{2}+\frac{\omega_{1}^{2}q_{1}^{2}}{2}+\frac{\omega_{2}^{2}q_{2}^{2}}{2}.
\end{align}
The exact result for the spectrum, which can be found analytically
as the product of two TA-SCIVR spectra,\cite{Ceotto_AspuruGuzik_Multiplecoherent_2009}
reads
\begin{align}
 & I_{\text{HK}}(E)=\exp\left[-\frac{p_{\text{eq,1}}^{2}}{2\hbar\omega_{1}}\right]\exp\left[-\frac{p_{2,0}^{2}}{2\hbar\omega_{2}}\right]\nonumber \\
 & \times\sum\limits _{n,m}\frac{1}{2^{n+m}n!m!}\left(\frac{p_{\text{eq,1}}^{2}}{\hbar\omega_{1}}\right)^{n}\left(\frac{p_{\text{eq,2}}^{2}}{\hbar\omega_{2}}\right)^{m}\delta\left(E-\hbar\omega_{1}\left[n+\frac{1}{2}\right]-\hbar\omega_{2}\left[m+\frac{1}{2}\right]\right),\label{eq:2HOs_HK}
\end{align}
where initial conditions $(\mathbf{p}_{\text{eq}},\mathbf{0})$ have
been chosen for simplicity and $\hbar$ has been set to unity. Using
M-TA-SCIVR according to \equationabb(\ref{eq:HYB_sep}) instead and
treating the HO with index ``1'' with HK while describing the HO
with index ``2'' on the TG level, \cite{Buchholz_Ceotto_MixedSC_2016}
the result looks very similar
\begin{align}
 & I_{\text{M}}(E)=\exp\left[-\frac{p_{\text{eq,1}}^{2}}{2\hbar\omega_{1}}\right]\exp\left[-\frac{p_{\text{eq,2}}^{2}}{\hbar\omega_{2}}\right]\nonumber \\
 & \times\sum\limits _{n,m}\frac{1}{2^{n+2m}n!(m!)^{2}}\left(\frac{p_{\text{eq,1}}^{2}}{\hbar\omega_{1}}\right)^{n}\left(\frac{p_{\text{eq,2}}^{2}}{\hbar\omega_{2}}\right)^{2m}\delta\left(E-\hbar\omega_{1}\left[n+\frac{1}{2}\right]-\hbar\omega_{2}\left[m+\frac{1}{2}\right]\right).\label{eq:2HOs_mixed}
\end{align}
With the mixed approach, all peaks positions of the TG coordinate
are reproduced exactly, while the peak weights are changed such that
overtones are suppressed. \cite{Buchholz_Ceotto_MixedSC_2016} While
this inherent suppression is a nice feature, it is not enough to completely
remove peaks from a very noisy spectrum, as we have seen in \figureabb\ref{fig:Morse-CL-big}. 

We will therefore apply the SAM-TA-SCIVR to this problem, using the
same phase space separation as before. As in the analytic results
above, we will set the initial position to zero for notational brevity,
and replace $p_{\text{\text{hk}}}(0)$ with $p_{1}$ as well as $\left(p_{\text{eq},\text{hk}},p_{\text{eq},\text{tg}}\right)$
with $\left(p_{\text{eq},1},p_{\text{eq,2}}\right)$ for the same
reason. After unfolding the modulus in \equationabb(\ref{eq:SAM_TA_SCIVR}),
the SAM formulation thus becomes 
\begin{align}
I(E)= & \ \frac{1}{2\pi\hbar}\frac{1}{\pi\hbar T}\int\text{d}p_{1}\int\text{d}q_{1}\text{Re}\Bigg\{\int_{0}^{T}\text{d}t_{1}\int_{t_{1}}^{\infty}\text{d}t_{2}\nonumber \\
 & \times\text{e}^{\text{i}\left[E\left(t_{2}-t_{1}\right)+\phi\left(t_{2}\right)-\phi\left(t_{1}\right)+S\left(t_{2}\right)-S\left(t_{1}\right)\right]/\hbar}\nonumber \\
 & \times\left\langle p_{\text{eq,1}},0\middle|p_{1}\left(t_{2}\right),q_{1}\left(t_{2}\right)\right\rangle \left\langle p_{1}\left(t_{1}\right),q_{1}\left(t_{1}\right)\middle|p_{\text{eq,1}},0\right\rangle \Bigg\},\label{eq:2HOs_SAM_1}
\end{align}
where $\left(p_{1}\left(t\right),q_{1}\left(t\right)\right)$ is the
classical trajectory from \equationabb(\ref{eq:2HOs_eom}). Using
the classical trajectory with $\mathbf{q}_{0}=\mathbf{0}$, the action
becomes
\begin{align}
S(t) & =\left(\frac{p_{1}^{2}}{2\omega_{1}}-\frac{1}{2}\omega_{1}q_{1}^{2}\right)\cos\omega_{1}t\sin\omega_{1}t-p_{1}q_{1}\text{sin}^{2}\omega_{1}t+\frac{p_{\text{eq,2}}^{2}}{2\omega_{2}}\cos\omega_{2}t\sin\omega_{2}t,\label{eq:2HOs_action}
\end{align}
and the prefactor phase is 
\begin{align}
\phi\left(t\right) & =-\frac{\hbar\left(\omega_{1}+\omega_{2}\right)}{2}t.
\end{align}
The phase space integration over the HK coordinates can be performed
analytically, and the remaining time integrand in \equationabb(\ref{eq:2HOs_SAM_1})
can be written in the form of a large exponential, 
\begin{align}
\exp\left\{ \frac{\text{i}}{\hbar}\left[E\left(t_{2}-t_{1}\right)+\phi(t_{2})-\phi(t_{1})\right]+E_{\text{hk}}\left(t_{1},t_{2}\right)+E_{\text{tg}}\left(t_{1},t_{2}\right)\right\}  & ,
\end{align}
where the term $E_{\text{hk}}\left(t_{1},t_{2}\right)$ denotes the
phase space integrated contribution from the HK DOF,\cite{Ceotto_AspuruGuzik_Multiplecoherent_2009,Buchholz_Ceotto_MixedSC_2016}
and $E_{\text{tg}}\left(t_{1},t_{2}\right)$ is the contribution of
the TG part, which consists only of (part of) the action. Taking results
from Refs.$\ $\onlinecite{Ceotto_AspuruGuzik_Multiplecoherent_2009}
and \onlinecite{Buchholz_Ceotto_MixedSC_2016} and changing the integration
variables to $\tau=t_{2}-t_{1}$ and $\tau'=t_{1}$, the HK term becomes
\begin{align}
E_{\text{hk}}\left(\tau,\tau'\right) & =\frac{p_{\text{eq},1}^{2}}{2\hbar\omega_{1}}\left(\text{e}^{-\text{i}\omega_{1}\tau}-1\right).
\end{align}
For the TG term, we replace the trigonometric functions by exponentials
to find
\begin{align}
E_{\text{tg}}\left(\tau,\tau'\right) & =\frac{1}{4\hbar}\frac{p_{\text{eq,2}}^{2}}{2\omega_{2}}\left[\text{e}^{2\text{i}\omega_{2}\left(\tau+\tau'\right)}-\text{e}^{-2\text{i}\omega_{2}\left(\tau+\tau'\right)}-\text{e}^{2\text{i}\omega_{2}\tau'}+\text{e}^{-2\text{i}\omega_{2}\tau'}\right]\label{eq:2HOs_Etg}
\end{align}
This makes the intermediate expression for the SAM spectrum 
\begin{align}
I(E)= & \ \frac{1}{\pi\hbar T}\text{e}^{-p_{\text{eq},1}^{2}/\left(2\hbar\omega_{1}\right)}\text{Re}\Bigg\{\int_{0}^{T}\text{d}\tau'\int_{0}^{\infty}\text{d}\tau\ \text{e}^{\text{i}\left[E-\hbar\left(\omega_{1}+\omega_{2}\right)/2\right]\tau/\hbar}\nonumber \\
 & \times\exp\left\{ \frac{p_{\text{eq,1}}^{2}}{2\hbar\omega_{1}}\text{e}^{-\text{i}\omega_{1}\tau}+\frac{1}{4}\frac{p_{\text{eq,2}}^{2}}{2\hbar\omega_{2}}\left[\text{e}^{2\text{i}\omega_{2}\left(\tau+\tau'\right)}-\text{e}^{-2\text{i}\omega_{2}\left(\tau+\tau'\right)}+\text{e}^{-2\text{i}\omega_{2}\tau'}-\text{e}^{2\text{i}\omega_{2}\tau'}\right]\right\} \Bigg\}.
\end{align}
We can now write the exponential in the last line as a product with
five factors and replace the lower exponentials by their power series
expansions\cite{Ceotto_AspuruGuzik_Multiplecoherent_2009,Buchholz_Ceotto_MixedSC_2016}
\begin{align}
I(E)= & \ \frac{1}{\pi\hbar T}\text{e}^{-p_{\text{eq,1}}^{2}/\left(2\hbar\omega_{1}\right)}\text{Re}\Bigg\{\int_{0}^{T}\text{d}\tau'\int_{0}^{\infty}\text{d}\tau\ \text{e}^{\text{i}\left[E-\hbar\left(\omega_{1}+\omega_{2}\right)/2\right]\tau/\hbar}\nonumber \\
 & \times\sum_{n,m,k,l,o}\frac{\left(-1\right)^{k+o}}{n!m!k!l!o!}\left(\frac{p_{\text{eq,1}}^{2}}{2\hbar\omega_{1}}\right)^{n}\left(\frac{1}{4}\frac{p_{\text{eq,2}}^{2}}{2\hbar\omega_{2}}\right)^{m+k+l+o}\text{e}^{-\text{i}\omega_{1}n\tau}\text{e}^{2\text{i}\omega_{2}m\left(\tau+\tau'\right)}\text{e}^{-2\text{i}\omega_{2}k\left(\tau+\tau'\right)}\text{e}^{-2\text{i}l\omega_{2}\tau'}\text{e}^{2\text{i}o\omega_{2}\tau'}\Bigg\}.
\end{align}
 The resulting fivefold sum collapses to a sum over three variables
in the limit $T\rightarrow\infty$ because the factor $1/T$ needs
to cancel after integration over $\tau'$ in order for the contribution
to survive, which is the case only for $k=o$ and $l=m$, and we end
up with
\begin{align}
I(E)= & \ \frac{1}{\pi\hbar}\text{e}^{-p_{\text{eq,1}}^{2}/\left(2\hbar\omega_{1}\right)}\text{Re}\Bigg\{\int_{0}^{\infty}\text{d}\tau\ \text{e}^{\text{i}\left[E-\hbar\left(\omega_{1}+\omega_{2}\right)/2\right]\tau/\hbar}\nonumber \\
 & \times\sum_{n,k,l}\frac{1}{n!\left(k!l!\right)^{2}}\left(\frac{p_{\text{eq,1}}^{2}}{2\hbar\omega_{1}}\right)^{n}\left(\frac{1}{4}\frac{p_{\text{eq,2}}^{2}}{2\hbar\omega_{2}}\right)^{2k+2l}\exp\left[-\text{i}\omega_{1}n\tau-2\text{i}\omega_{2}\left(k-l\right)\tau\right]\Bigg\}.
\end{align}
After performing the remaining integration over $\tau$, the final
SAM spectrum for two uncoupled HOs emerges as
\begin{align}
 & I_{\text{SAM}}(E)=\exp\left[-\frac{p_{\text{eq,1}}^{2}}{2\hbar\omega_{1}}\right]\nonumber \\
 & \times\sum\limits _{n,k,l}\frac{1}{2^{n}2^{3(2k+2l)}}\frac{1}{n!(k!)^{2}(l!)^{2}}\left(\frac{p_{\text{eq,1}}^{2}}{\hbar\omega_{1}}\right)^{n}\left(\frac{p_{\text{eq,2}}^{2}}{\hbar\omega_{2}}\right)^{2k+2l}\delta\left(E-\hbar\omega_{1}\left[n+\frac{1}{2}\right]-\hbar\omega_{2}\left[2(k-l)+\frac{1}{2}\right]\right)\label{eq:2HOs_SAM_fin}
\end{align}
From the comparison to the exact and mixed results in \equationsabb(\ref{eq:2HOs_HK})
and (\ref{eq:2HOs_mixed}), we get an insight into the nature of this
approximation. First, setting $n=k=l=0$, the correct ground state
energy of the composed system is recovered. Second, by comparing the
case $k=l=0$ in \equationabb(\ref{eq:2HOs_SAM_fin}) to $m=0$ in
the other expressions, we find all three spectra for the HK DOF to
be exactly the same. Third, the SAM expression turns out to contain
only even peaks in the TG coordinate. This is an important property
given that these DOFs are usually centered around the ground state
energy initially, which means that the first excitations are the main
source of unwanted, noisy peaks. Finally, we see that all even TG
excitations have an unphysical counterpart at the respective ``negative
frequency''. The correct second excitation at $\omega_{1}+5\omega_{2}/2$
($n=l=0$, $k=1$), for example, has a ghost peak equivalent at $\omega_{1}-5\omega_{2}/2$
($n=k=0$, $l=1$). However, all TG peaks are systematically much
smaller than the closest HK peaks. This is why we think that the ghost
peaks are a small price to pay for a cheap built-in filter that removes
odd TG excitations and thus makes spectra accessible that would be
just noise otherwise. In the numerical examples in \sectionabb\ref{sec:results},
we show that these properties hold approximately also for anharmonic,
coupled systems.

\newpage{}

\bibliographystyle{apsrev4-1}
\bibliography{SEMICOMPLEX}

\begin{thebibliography}{127}%
\makeatletter
\providecommand \@ifxundefined [1]{%
 \@ifx{#1\undefined}
}%
\providecommand \@ifnum [1]{%
 \ifnum #1\expandafter \@firstoftwo
 \else \expandafter \@secondoftwo
 \fi
}%
\providecommand \@ifx [1]{%
 \ifx #1\expandafter \@firstoftwo
 \else \expandafter \@secondoftwo
 \fi
}%
\providecommand \natexlab [1]{#1}%
\providecommand \enquote  [1]{``#1''}%
\providecommand \bibnamefont  [1]{#1}%
\providecommand \bibfnamefont [1]{#1}%
\providecommand \citenamefont [1]{#1}%
\providecommand \href@noop [0]{\@secondoftwo}%
\providecommand \href [0]{\begingroup \@sanitize@url \@href}%
\providecommand \@href[1]{\@@startlink{#1}\@@href}%
\providecommand \@@href[1]{\endgroup#1\@@endlink}%
\providecommand \@sanitize@url [0]{\catcode `\\12\catcode `\$12\catcode
  `\&12\catcode `\#12\catcode `\^12\catcode `\_12\catcode `\%12\relax}%
\providecommand \@@startlink[1]{}%
\providecommand \@@endlink[0]{}%
\providecommand \url  [0]{\begingroup\@sanitize@url \@url }%
\providecommand \@url [1]{\endgroup\@href {#1}{\urlprefix }}%
\providecommand \urlprefix  [0]{URL }%
\providecommand \Eprint [0]{\href }%
\providecommand \doibase [0]{http://dx.doi.org/}%
\providecommand \selectlanguage [0]{\@gobble}%
\providecommand \bibinfo  [0]{\@secondoftwo}%
\providecommand \bibfield  [0]{\@secondoftwo}%
\providecommand \translation [1]{[#1]}%
\providecommand \BibitemOpen [0]{}%
\providecommand \bibitemStop [0]{}%
\providecommand \bibitemNoStop [0]{.\EOS\space}%
\providecommand \EOS [0]{\spacefactor3000\relax}%
\providecommand \BibitemShut  [1]{\csname bibitem#1\endcsname}%
\let\auto@bib@innerbib\@empty
\bibitem [{\citenamefont {Miller}(1968)}]{miller1968uniform}%
  \BibitemOpen
  \bibfield  {author} {\bibinfo {author} {\bibfnamefont {W.~H.}\ \bibnamefont
  {Miller}},\ }\href@noop {} {\bibfield  {journal} {\bibinfo  {journal} {J.
  Chem. Phys.}\ }\textbf {\bibinfo {volume} {48}},\ \bibinfo {pages} {464}
  (\bibinfo {year} {1968})}\BibitemShut {NoStop}%
\bibitem [{\citenamefont {{Wolken Jr}}\ \emph {et~al.}(1972)\citenamefont
  {{Wolken Jr}}, \citenamefont {Miller},\ and\ \citenamefont
  {Karplus}}]{wolken1972theoretical}%
  \BibitemOpen
  \bibfield  {author} {\bibinfo {author} {\bibfnamefont {G.}~\bibnamefont
  {{Wolken Jr}}}, \bibinfo {author} {\bibfnamefont {W.~H.}\ \bibnamefont
  {Miller}}, \ and\ \bibinfo {author} {\bibfnamefont {M.}~\bibnamefont
  {Karplus}},\ }\href@noop {} {\bibfield  {journal} {\bibinfo  {journal} {J.
  Chem. Phys.}\ }\textbf {\bibinfo {volume} {56}},\ \bibinfo {pages} {4930}
  (\bibinfo {year} {1972})}\BibitemShut {NoStop}%
\bibitem [{\citenamefont {Garrett}\ and\ \citenamefont
  {Miller}(1978)}]{garrett1978quantum}%
  \BibitemOpen
  \bibfield  {author} {\bibinfo {author} {\bibfnamefont {B.~C.}\ \bibnamefont
  {Garrett}}\ and\ \bibinfo {author} {\bibfnamefont {W.~H.}\ \bibnamefont
  {Miller}},\ }\href@noop {} {\bibfield  {journal} {\bibinfo  {journal} {J.
  Chem. Phys.}\ }\textbf {\bibinfo {volume} {68}},\ \bibinfo {pages} {4051}
  (\bibinfo {year} {1978})}\BibitemShut {NoStop}%
\bibitem [{\citenamefont {Miller}(1971)}]{miller1971semiclassical}%
  \BibitemOpen
  \bibfield  {author} {\bibinfo {author} {\bibfnamefont {W.~H.}\ \bibnamefont
  {Miller}},\ }\href@noop {} {\bibfield  {journal} {\bibinfo  {journal}
  {Accounts of Chemical Research}\ }\textbf {\bibinfo {volume} {4}},\ \bibinfo
  {pages} {161} (\bibinfo {year} {1971})}\BibitemShut {NoStop}%
\bibitem [{\citenamefont {Bowman}\ and\ \citenamefont
  {Lee}(1980)}]{bowman1980sudden}%
  \BibitemOpen
  \bibfield  {author} {\bibinfo {author} {\bibfnamefont {J.~M.}\ \bibnamefont
  {Bowman}}\ and\ \bibinfo {author} {\bibfnamefont {K.~T.}\ \bibnamefont
  {Lee}},\ }\href@noop {} {\bibfield  {journal} {\bibinfo  {journal} {J. Chem.
  Phys.}\ }\textbf {\bibinfo {volume} {72}},\ \bibinfo {pages} {5071} (\bibinfo
  {year} {1980})}\BibitemShut {NoStop}%
\bibitem [{\citenamefont {Bowman}\ and\ \citenamefont
  {Kupperman}(1973)}]{bowman1973semi}%
  \BibitemOpen
  \bibfield  {author} {\bibinfo {author} {\bibfnamefont {J.}~\bibnamefont
  {Bowman}}\ and\ \bibinfo {author} {\bibfnamefont {A.}~\bibnamefont
  {Kupperman}},\ }\href@noop {} {\bibfield  {journal} {\bibinfo  {journal}
  {Chem. Phys.}\ }\textbf {\bibinfo {volume} {2}},\ \bibinfo {pages} {158}
  (\bibinfo {year} {1973})}\BibitemShut {NoStop}%
\bibitem [{\citenamefont {Schatz}\ \emph {et~al.}(1975)\citenamefont {Schatz},
  \citenamefont {Bowman},\ and\ \citenamefont {Kuppermann}}]{schatz1975exact}%
  \BibitemOpen
  \bibfield  {author} {\bibinfo {author} {\bibfnamefont {G.~C.}\ \bibnamefont
  {Schatz}}, \bibinfo {author} {\bibfnamefont {J.~M.}\ \bibnamefont {Bowman}},
  \ and\ \bibinfo {author} {\bibfnamefont {A.}~\bibnamefont {Kuppermann}},\
  }\href@noop {} {\bibfield  {journal} {\bibinfo  {journal} {J. Chem. Phys.}\
  }\textbf {\bibinfo {volume} {63}},\ \bibinfo {pages} {674} (\bibinfo {year}
  {1975})}\BibitemShut {NoStop}%
\bibitem [{\citenamefont {Gianturco}(1976)}]{gianturco1976scattering}%
  \BibitemOpen
  \bibfield  {author} {\bibinfo {author} {\bibfnamefont {F.}~\bibnamefont
  {Gianturco}},\ }\href@noop {} {\bibfield  {journal} {\bibinfo  {journal}
  {International Journal of Quantum Chemistry}\ }\textbf {\bibinfo {volume}
  {10}},\ \bibinfo {pages} {37} (\bibinfo {year} {1976})}\BibitemShut {NoStop}%
\bibitem [{\citenamefont {Aquilanti}\ and\ \citenamefont
  {Casavecchia}(1976)}]{aquilanti1976total}%
  \BibitemOpen
  \bibfield  {author} {\bibinfo {author} {\bibfnamefont {V.}~\bibnamefont
  {Aquilanti}}\ and\ \bibinfo {author} {\bibfnamefont {P.}~\bibnamefont
  {Casavecchia}},\ }\href@noop {} {\bibfield  {journal} {\bibinfo  {journal}
  {J. Chem. Phys.}\ }\textbf {\bibinfo {volume} {64}},\ \bibinfo {pages} {751}
  (\bibinfo {year} {1976})}\BibitemShut {NoStop}%
\bibitem [{\citenamefont {Aquilanti}\ and\ \citenamefont
  {Lagan{\`a}}(1975)}]{aquilanti1975resonant}%
  \BibitemOpen
  \bibfield  {author} {\bibinfo {author} {\bibfnamefont {V.}~\bibnamefont
  {Aquilanti}}\ and\ \bibinfo {author} {\bibfnamefont {A.}~\bibnamefont
  {Lagan{\`a}}},\ }\href@noop {} {\bibfield  {journal} {\bibinfo  {journal}
  {Zeitschrift f{\"u}r Physikalische Chemie}\ }\textbf {\bibinfo {volume}
  {96}},\ \bibinfo {pages} {229} (\bibinfo {year} {1975})}\BibitemShut
  {NoStop}%
\bibitem [{\citenamefont {Bonnet}\ and\ \citenamefont
  {Rayez}(1997)}]{bonnet1997quasiclassical}%
  \BibitemOpen
  \bibfield  {author} {\bibinfo {author} {\bibfnamefont {L.}~\bibnamefont
  {Bonnet}}\ and\ \bibinfo {author} {\bibfnamefont {J.}~\bibnamefont {Rayez}},\
  }\href@noop {} {\bibfield  {journal} {\bibinfo  {journal} {Chem. Phys.
  Lett.}\ }\textbf {\bibinfo {volume} {277}},\ \bibinfo {pages} {183} (\bibinfo
  {year} {1997})}\BibitemShut {NoStop}%
\bibitem [{\citenamefont {Aquilanti}\ \emph {et~al.}(2000)\citenamefont
  {Aquilanti}, \citenamefont {Beddoni}, \citenamefont {Cavalli}, \citenamefont
  {Lombardi},\ and\ \citenamefont {Littlejohn}}]{aquilanti2000collective}%
  \BibitemOpen
  \bibfield  {author} {\bibinfo {author} {\bibfnamefont {V.}~\bibnamefont
  {Aquilanti}}, \bibinfo {author} {\bibfnamefont {A.}~\bibnamefont {Beddoni}},
  \bibinfo {author} {\bibfnamefont {S.}~\bibnamefont {Cavalli}}, \bibinfo
  {author} {\bibfnamefont {A.}~\bibnamefont {Lombardi}}, \ and\ \bibinfo
  {author} {\bibfnamefont {R.}~\bibnamefont {Littlejohn}},\ }\href@noop {}
  {\bibfield  {journal} {\bibinfo  {journal} {Molecular Physics}\ }\textbf
  {\bibinfo {volume} {98}},\ \bibinfo {pages} {1763} (\bibinfo {year}
  {2000})}\BibitemShut {NoStop}%
\bibitem [{\citenamefont {Connor}\ \emph {et~al.}(1979)\citenamefont {Connor},
  \citenamefont {Jakubetz},\ and\ \citenamefont
  {Lagana}}]{connor1979comparison}%
  \BibitemOpen
  \bibfield  {author} {\bibinfo {author} {\bibfnamefont {J.}~\bibnamefont
  {Connor}}, \bibinfo {author} {\bibfnamefont {W.}~\bibnamefont {Jakubetz}}, \
  and\ \bibinfo {author} {\bibfnamefont {A.}~\bibnamefont {Lagana}},\
  }\href@noop {} {\bibfield  {journal} {\bibinfo  {journal} {Journal of
  Physical Chemistry}\ }\textbf {\bibinfo {volume} {83}},\ \bibinfo {pages}
  {73} (\bibinfo {year} {1979})}\BibitemShut {NoStop}%
\bibitem [{\citenamefont {{De Fazio}}\ \emph {et~al.}(1994)\citenamefont {{De
  Fazio}}, \citenamefont {Gianturco}, \citenamefont {Rodriguez-Ruiz},
  \citenamefont {Tang},\ and\ \citenamefont {Toennies}}]{de1994semiclassical}%
  \BibitemOpen
  \bibfield  {author} {\bibinfo {author} {\bibfnamefont {D.}~\bibnamefont {{De
  Fazio}}}, \bibinfo {author} {\bibfnamefont {F.}~\bibnamefont {Gianturco}},
  \bibinfo {author} {\bibfnamefont {J.}~\bibnamefont {Rodriguez-Ruiz}},
  \bibinfo {author} {\bibfnamefont {K.}~\bibnamefont {Tang}}, \ and\ \bibinfo
  {author} {\bibfnamefont {J.}~\bibnamefont {Toennies}},\ }\href@noop {}
  {\bibfield  {journal} {\bibinfo  {journal} {Journal of Physics B: Atomic,
  Molecular and Optical Physics}\ }\textbf {\bibinfo {volume} {27}},\ \bibinfo
  {pages} {303} (\bibinfo {year} {1994})}\BibitemShut {NoStop}%
\bibitem [{\citenamefont {Miller}\ \emph {et~al.}(2003)\citenamefont {Miller},
  \citenamefont {Zhao}, \citenamefont {Ceotto},\ and\ \citenamefont
  {Yang}}]{Miller_Ceotto_2003QI}%
  \BibitemOpen
  \bibfield  {author} {\bibinfo {author} {\bibfnamefont {W.~H.}\ \bibnamefont
  {Miller}}, \bibinfo {author} {\bibfnamefont {Y.}~\bibnamefont {Zhao}},
  \bibinfo {author} {\bibfnamefont {M.}~\bibnamefont {Ceotto}}, \ and\ \bibinfo
  {author} {\bibfnamefont {S.}~\bibnamefont {Yang}},\ }\href@noop {} {\bibfield
   {journal} {\bibinfo  {journal} {J. Chem. Phys.}\ }\textbf {\bibinfo {volume}
  {119}},\ \bibinfo {pages} {1329} (\bibinfo {year} {2003})}\BibitemShut
  {NoStop}%
\bibitem [{\citenamefont {Ceotto}\ and\ \citenamefont
  {Miller}(2004)}]{Ceotto_Miller_2004test}%
  \BibitemOpen
  \bibfield  {author} {\bibinfo {author} {\bibfnamefont {M.}~\bibnamefont
  {Ceotto}}\ and\ \bibinfo {author} {\bibfnamefont {W.~H.}\ \bibnamefont
  {Miller}},\ }\href@noop {} {\bibfield  {journal} {\bibinfo  {journal} {J.
  Chem. Phys.}\ }\textbf {\bibinfo {volume} {120}},\ \bibinfo {pages} {6356}
  (\bibinfo {year} {2004})}\BibitemShut {NoStop}%
\bibitem [{\citenamefont {Ceotto}\ \emph {et~al.}(2005)\citenamefont {Ceotto},
  \citenamefont {Yang},\ and\ \citenamefont
  {Miller}}]{Ceotto_Yang_2005quantum}%
  \BibitemOpen
  \bibfield  {author} {\bibinfo {author} {\bibfnamefont {M.}~\bibnamefont
  {Ceotto}}, \bibinfo {author} {\bibfnamefont {S.}~\bibnamefont {Yang}}, \ and\
  \bibinfo {author} {\bibfnamefont {W.~H.}\ \bibnamefont {Miller}},\
  }\href@noop {} {\bibfield  {journal} {\bibinfo  {journal} {J. Chem. Phys.}\
  }\textbf {\bibinfo {volume} {122}},\ \bibinfo {pages} {044109} (\bibinfo
  {year} {2005})}\BibitemShut {NoStop}%
\bibitem [{\citenamefont {Aieta}\ and\ \citenamefont
  {Ceotto}(2017)}]{Ceotto_Aieta2017quantumTST}%
  \BibitemOpen
  \bibfield  {author} {\bibinfo {author} {\bibfnamefont {C.}~\bibnamefont
  {Aieta}}\ and\ \bibinfo {author} {\bibfnamefont {M.}~\bibnamefont {Ceotto}},\
  }\href@noop {} {\bibfield  {journal} {\bibinfo  {journal} {J. Chem. Phys.}\
  }\textbf {\bibinfo {volume} {146}},\ \bibinfo {pages} {214115} (\bibinfo
  {year} {2017})}\BibitemShut {NoStop}%
\bibitem [{\citenamefont {Aieta}\ \emph {et~al.}(2016)\citenamefont {Aieta},
  \citenamefont {Gabas},\ and\ \citenamefont {Ceotto}}]{Ceotto_Aieta_Gabas_16}%
  \BibitemOpen
  \bibfield  {author} {\bibinfo {author} {\bibfnamefont {C.}~\bibnamefont
  {Aieta}}, \bibinfo {author} {\bibfnamefont {F.}~\bibnamefont {Gabas}}, \ and\
  \bibinfo {author} {\bibfnamefont {M.}~\bibnamefont {Ceotto}},\ }\href@noop {}
  {\bibfield  {journal} {\bibinfo  {journal} {J. Phys. Chem. A}\ }\textbf
  {\bibinfo {volume} {120}},\ \bibinfo {pages} {4853} (\bibinfo {year}
  {2016})}\BibitemShut {NoStop}%
\bibitem [{\citenamefont {Mandr{\`a}}\ \emph {et~al.}(2013)\citenamefont
  {Mandr{\`a}}, \citenamefont {Valleau},\ and\ \citenamefont
  {Ceotto}}]{Ceotto_Mandra2013deep}%
  \BibitemOpen
  \bibfield  {author} {\bibinfo {author} {\bibfnamefont {S.}~\bibnamefont
  {Mandr{\`a}}}, \bibinfo {author} {\bibfnamefont {S.}~\bibnamefont {Valleau}},
  \ and\ \bibinfo {author} {\bibfnamefont {M.}~\bibnamefont {Ceotto}},\
  }\href@noop {} {\bibfield  {journal} {\bibinfo  {journal} {International
  Journal of Quantum Chemistry}\ }\textbf {\bibinfo {volume} {113}},\ \bibinfo
  {pages} {1722} (\bibinfo {year} {2013})}\BibitemShut {NoStop}%
\bibitem [{\citenamefont {Mandr{\`a}}\ \emph {et~al.}(2014)\citenamefont
  {Mandr{\`a}}, \citenamefont {Schrier},\ and\ \citenamefont
  {Ceotto}}]{Ceotto_Mandra2014helium}%
  \BibitemOpen
  \bibfield  {author} {\bibinfo {author} {\bibfnamefont {S.}~\bibnamefont
  {Mandr{\`a}}}, \bibinfo {author} {\bibfnamefont {J.}~\bibnamefont {Schrier}},
  \ and\ \bibinfo {author} {\bibfnamefont {M.}~\bibnamefont {Ceotto}},\
  }\href@noop {} {\bibfield  {journal} {\bibinfo  {journal} {J. Phys. Chem. A}\
  }\textbf {\bibinfo {volume} {118}},\ \bibinfo {pages} {6457} (\bibinfo {year}
  {2014})}\BibitemShut {NoStop}%
\bibitem [{\citenamefont {Conte}\ \emph
  {et~al.}(2015{\natexlab{a}})\citenamefont {Conte}, \citenamefont {Houston},\
  and\ \citenamefont {Bowman}}]{Conte_Bowman_CollisionsCH4-H2O_2015}%
  \BibitemOpen
  \bibfield  {author} {\bibinfo {author} {\bibfnamefont {R.}~\bibnamefont
  {Conte}}, \bibinfo {author} {\bibfnamefont {P.~L.}\ \bibnamefont {Houston}},
  \ and\ \bibinfo {author} {\bibfnamefont {J.~M.}\ \bibnamefont {Bowman}},\
  }\href@noop {} {\bibfield  {journal} {\bibinfo  {journal} {J. Phys. Chem. A}\
  }\textbf {\bibinfo {volume} {119}},\ \bibinfo {pages} {12304} (\bibinfo
  {year} {2015}{\natexlab{a}})}\BibitemShut {NoStop}%
\bibitem [{\citenamefont {Conte}\ \emph
  {et~al.}(2015{\natexlab{b}})\citenamefont {Conte}, \citenamefont {Qu},\ and\
  \citenamefont {Bowman}}]{Conte_Bowman_Manybody_2015}%
  \BibitemOpen
  \bibfield  {author} {\bibinfo {author} {\bibfnamefont {R.}~\bibnamefont
  {Conte}}, \bibinfo {author} {\bibfnamefont {C.}~\bibnamefont {Qu}}, \ and\
  \bibinfo {author} {\bibfnamefont {J.~M.}\ \bibnamefont {Bowman}},\
  }\href@noop {} {\bibfield  {journal} {\bibinfo  {journal} {J. Chem. Theory
  Comput.}\ }\textbf {\bibinfo {volume} {11}},\ \bibinfo {pages} {1631}
  (\bibinfo {year} {2015}{\natexlab{b}})}\BibitemShut {NoStop}%
\bibitem [{\citenamefont {Homayoon}\ \emph {et~al.}(2015)\citenamefont
  {Homayoon}, \citenamefont {Conte}, \citenamefont {Qu},\ and\ \citenamefont
  {Bowman}}]{Homayoon_Bowman_H2-H2O_2015}%
  \BibitemOpen
  \bibfield  {author} {\bibinfo {author} {\bibfnamefont {Z.}~\bibnamefont
  {Homayoon}}, \bibinfo {author} {\bibfnamefont {R.}~\bibnamefont {Conte}},
  \bibinfo {author} {\bibfnamefont {C.}~\bibnamefont {Qu}}, \ and\ \bibinfo
  {author} {\bibfnamefont {J.~M.}\ \bibnamefont {Bowman}},\ }\href@noop {}
  {\bibfield  {journal} {\bibinfo  {journal} {J. Chem. Phys.}\ }\textbf
  {\bibinfo {volume} {143}},\ \bibinfo {pages} {084302} (\bibinfo {year}
  {2015})}\BibitemShut {NoStop}%
\bibitem [{\citenamefont {Houston}\ \emph {et~al.}(2016)\citenamefont
  {Houston}, \citenamefont {Conte},\ and\ \citenamefont
  {Bowman}}]{Houston_Bowman_RoamingH2CO_2016}%
  \BibitemOpen
  \bibfield  {author} {\bibinfo {author} {\bibfnamefont {P.~L.}\ \bibnamefont
  {Houston}}, \bibinfo {author} {\bibfnamefont {R.}~\bibnamefont {Conte}}, \
  and\ \bibinfo {author} {\bibfnamefont {J.~M.}\ \bibnamefont {Bowman}},\
  }\href@noop {} {\bibfield  {journal} {\bibinfo  {journal} {J. Phys. Chem. A}\
  }\textbf {\bibinfo {volume} {120}},\ \bibinfo {pages} {5103} (\bibinfo {year}
  {2016})}\BibitemShut {NoStop}%
\bibitem [{\citenamefont {Qu}\ \emph {et~al.}(2015)\citenamefont {Qu},
  \citenamefont {Conte}, \citenamefont {Houston},\ and\ \citenamefont
  {Bowman}}]{Chen_Bowman_Methane-Water_2015}%
  \BibitemOpen
  \bibfield  {author} {\bibinfo {author} {\bibfnamefont {C.}~\bibnamefont
  {Qu}}, \bibinfo {author} {\bibfnamefont {R.}~\bibnamefont {Conte}}, \bibinfo
  {author} {\bibfnamefont {P.~L.}\ \bibnamefont {Houston}}, \ and\ \bibinfo
  {author} {\bibfnamefont {J.~M.}\ \bibnamefont {Bowman}},\ }\href@noop {}
  {\bibfield  {journal} {\bibinfo  {journal} {Phys. Chem. Chem. Phys.}\
  }\textbf {\bibinfo {volume} {17}},\ \bibinfo {pages} {8172} (\bibinfo {year}
  {2015})}\BibitemShut {NoStop}%
\bibitem [{\citenamefont {Conte}\ \emph {et~al.}(2014)\citenamefont {Conte},
  \citenamefont {Houston},\ and\ \citenamefont
  {Bowman}}]{Conte_Bowman_Communication_2014}%
  \BibitemOpen
  \bibfield  {author} {\bibinfo {author} {\bibfnamefont {R.}~\bibnamefont
  {Conte}}, \bibinfo {author} {\bibfnamefont {P.~L.}\ \bibnamefont {Houston}},
  \ and\ \bibinfo {author} {\bibfnamefont {J.~M.}\ \bibnamefont {Bowman}},\
  }\href@noop {} {\bibfield  {journal} {\bibinfo  {journal} {J. Chem. Phys.}\
  }\textbf {\bibinfo {volume} {140}},\ \bibinfo {pages} {151101} (\bibinfo
  {year} {2014})}\BibitemShut {NoStop}%
\bibitem [{\citenamefont {Conte}\ \emph
  {et~al.}(2013{\natexlab{a}})\citenamefont {Conte}, \citenamefont {Houston},\
  and\ \citenamefont {Bowman}}]{Conte_Houston_Bowman_Ar2013}%
  \BibitemOpen
  \bibfield  {author} {\bibinfo {author} {\bibfnamefont {R.}~\bibnamefont
  {Conte}}, \bibinfo {author} {\bibfnamefont {P.~L.}\ \bibnamefont {Houston}},
  \ and\ \bibinfo {author} {\bibfnamefont {J.~M.}\ \bibnamefont {Bowman}},\
  }\href@noop {} {\bibfield  {journal} {\bibinfo  {journal} {J. Phys. Chem. A}\
  }\textbf {\bibinfo {volume} {117}},\ \bibinfo {pages} {14028} (\bibinfo
  {year} {2013}{\natexlab{a}})}\BibitemShut {NoStop}%
\bibitem [{\citenamefont {Feynman}\ and\ \citenamefont
  {Hibbs}(1965)}]{feynman_pathintegral_1965}%
  \BibitemOpen
  \bibfield  {author} {\bibinfo {author} {\bibfnamefont {R.~P.}\ \bibnamefont
  {Feynman}}\ and\ \bibinfo {author} {\bibfnamefont {A.~R.}\ \bibnamefont
  {Hibbs}},\ }\href@noop {} {\emph {\bibinfo {title} {{Quantum mechanics and
  path integrals}}}}\ (\bibinfo  {publisher} {McGraw-Hill},\ \bibinfo {year}
  {1965})\BibitemShut {NoStop}%
\bibitem [{\citenamefont {Ceperley}(1995)}]{Ceperley1994_reviewPIMC}%
  \BibitemOpen
  \bibfield  {author} {\bibinfo {author} {\bibfnamefont {D.~M.}\ \bibnamefont
  {Ceperley}},\ }\href {\doibase 10.1103/RevModPhys.67.279} {\bibfield
  {journal} {\bibinfo  {journal} {Rev. Mod. Phys.}\ }\textbf {\bibinfo {volume}
  {67}},\ \bibinfo {pages} {279} (\bibinfo {year} {1995})}\BibitemShut
  {NoStop}%
\bibitem [{\citenamefont {Parrinello}\ and\ \citenamefont
  {Rahman}(1984)}]{parrinello1984study}%
  \BibitemOpen
  \bibfield  {author} {\bibinfo {author} {\bibfnamefont {M.}~\bibnamefont
  {Parrinello}}\ and\ \bibinfo {author} {\bibfnamefont {A.}~\bibnamefont
  {Rahman}},\ }\href@noop {} {\bibfield  {journal} {\bibinfo  {journal} {J.
  Chem. Phys.}\ }\textbf {\bibinfo {volume} {80}},\ \bibinfo {pages} {860}
  (\bibinfo {year} {1984})}\BibitemShut {NoStop}%
\bibitem [{\citenamefont {Marx}\ and\ \citenamefont
  {Parrinello}(1996)}]{marx1996ab}%
  \BibitemOpen
  \bibfield  {author} {\bibinfo {author} {\bibfnamefont {D.}~\bibnamefont
  {Marx}}\ and\ \bibinfo {author} {\bibfnamefont {M.}~\bibnamefont
  {Parrinello}},\ }\href@noop {} {\bibfield  {journal} {\bibinfo  {journal} {J.
  Chem. Phys.}\ }\textbf {\bibinfo {volume} {104}},\ \bibinfo {pages} {4077}
  (\bibinfo {year} {1996})}\BibitemShut {NoStop}%
\bibitem [{\citenamefont {Tuckerman}\ \emph {et~al.}(1996)\citenamefont
  {Tuckerman}, \citenamefont {Marx}, \citenamefont {Klein},\ and\ \citenamefont
  {Parrinello}}]{tuckerman1996efficient}%
  \BibitemOpen
  \bibfield  {author} {\bibinfo {author} {\bibfnamefont {M.~E.}\ \bibnamefont
  {Tuckerman}}, \bibinfo {author} {\bibfnamefont {D.}~\bibnamefont {Marx}},
  \bibinfo {author} {\bibfnamefont {M.~L.}\ \bibnamefont {Klein}}, \ and\
  \bibinfo {author} {\bibfnamefont {M.}~\bibnamefont {Parrinello}},\
  }\href@noop {} {\bibfield  {journal} {\bibinfo  {journal} {J. Chem. Phys.}\
  }\textbf {\bibinfo {volume} {104}},\ \bibinfo {pages} {5579} (\bibinfo {year}
  {1996})}\BibitemShut {NoStop}%
\bibitem [{\citenamefont {Cao}\ and\ \citenamefont
  {Voth}(1994)}]{cao1994formulation}%
  \BibitemOpen
  \bibfield  {author} {\bibinfo {author} {\bibfnamefont {J.}~\bibnamefont
  {Cao}}\ and\ \bibinfo {author} {\bibfnamefont {G.~A.}\ \bibnamefont {Voth}},\
  }\href@noop {} {\bibfield  {journal} {\bibinfo  {journal} {J. Chem. Phys.}\
  }\textbf {\bibinfo {volume} {100}},\ \bibinfo {pages} {5093} (\bibinfo {year}
  {1994})}\BibitemShut {NoStop}%
\bibitem [{\citenamefont {Geva}\ \emph {et~al.}(2001)\citenamefont {Geva},
  \citenamefont {Shi},\ and\ \citenamefont {Voth}}]{geva2001quantum}%
  \BibitemOpen
  \bibfield  {author} {\bibinfo {author} {\bibfnamefont {E.}~\bibnamefont
  {Geva}}, \bibinfo {author} {\bibfnamefont {Q.}~\bibnamefont {Shi}}, \ and\
  \bibinfo {author} {\bibfnamefont {G.~A.}\ \bibnamefont {Voth}},\ }\href@noop
  {} {\bibfield  {journal} {\bibinfo  {journal} {J. Chem. Phys.}\ }\textbf
  {\bibinfo {volume} {115}},\ \bibinfo {pages} {9209} (\bibinfo {year}
  {2001})}\BibitemShut {NoStop}%
\bibitem [{\citenamefont {Paesani}\ \emph {et~al.}(2009)\citenamefont
  {Paesani}, \citenamefont {Xantheas},\ and\ \citenamefont
  {Voth}}]{paesani2009infrared}%
  \BibitemOpen
  \bibfield  {author} {\bibinfo {author} {\bibfnamefont {F.}~\bibnamefont
  {Paesani}}, \bibinfo {author} {\bibfnamefont {S.~S.}\ \bibnamefont
  {Xantheas}}, \ and\ \bibinfo {author} {\bibfnamefont {G.~A.}\ \bibnamefont
  {Voth}},\ }\href@noop {} {\bibfield  {journal} {\bibinfo  {journal} {J. Phys.
  Chem. B}\ }\textbf {\bibinfo {volume} {113}},\ \bibinfo {pages} {13118}
  (\bibinfo {year} {2009})}\BibitemShut {NoStop}%
\bibitem [{\citenamefont {Poulsen}\ and\ \citenamefont
  {Rossky}(2001)}]{poulsen2001path}%
  \BibitemOpen
  \bibfield  {author} {\bibinfo {author} {\bibfnamefont {J.~A.}\ \bibnamefont
  {Poulsen}}\ and\ \bibinfo {author} {\bibfnamefont {P.~J.}\ \bibnamefont
  {Rossky}},\ }\href@noop {} {\bibfield  {journal} {\bibinfo  {journal} {J.
  Chem. Phys.}\ }\textbf {\bibinfo {volume} {115}},\ \bibinfo {pages} {8024}
  (\bibinfo {year} {2001})}\BibitemShut {NoStop}%
\bibitem [{\citenamefont {Habershon}\ \emph {et~al.}(2013)\citenamefont
  {Habershon}, \citenamefont {Manolopoulos}, \citenamefont {Markland},\ and\
  \citenamefont {{Miller III}}}]{habershon2013ring}%
  \BibitemOpen
  \bibfield  {author} {\bibinfo {author} {\bibfnamefont {S.}~\bibnamefont
  {Habershon}}, \bibinfo {author} {\bibfnamefont {D.~E.}\ \bibnamefont
  {Manolopoulos}}, \bibinfo {author} {\bibfnamefont {T.~E.}\ \bibnamefont
  {Markland}}, \ and\ \bibinfo {author} {\bibfnamefont {T.~F.}\ \bibnamefont
  {{Miller III}}},\ }\href@noop {} {\bibfield  {journal} {\bibinfo  {journal}
  {Annu. Rev. Phys. Chem.}\ }\textbf {\bibinfo {volume} {64}},\ \bibinfo
  {pages} {387} (\bibinfo {year} {2013})}\BibitemShut {NoStop}%
\bibitem [{\citenamefont {Craig}\ and\ \citenamefont
  {Manolopoulos}(2005{\natexlab{a}})}]{craig2005chemical}%
  \BibitemOpen
  \bibfield  {author} {\bibinfo {author} {\bibfnamefont {I.~R.}\ \bibnamefont
  {Craig}}\ and\ \bibinfo {author} {\bibfnamefont {D.~E.}\ \bibnamefont
  {Manolopoulos}},\ }\href@noop {} {\bibfield  {journal} {\bibinfo  {journal}
  {J. Chem. Phys.}\ }\textbf {\bibinfo {volume} {122}},\ \bibinfo {pages}
  {084106} (\bibinfo {year} {2005}{\natexlab{a}})}\BibitemShut {NoStop}%
\bibitem [{\citenamefont {Craig}\ and\ \citenamefont
  {Manolopoulos}(2005{\natexlab{b}})}]{craig2005refined}%
  \BibitemOpen
  \bibfield  {author} {\bibinfo {author} {\bibfnamefont {I.~R.}\ \bibnamefont
  {Craig}}\ and\ \bibinfo {author} {\bibfnamefont {D.~E.}\ \bibnamefont
  {Manolopoulos}},\ }\href@noop {} {\bibfield  {journal} {\bibinfo  {journal}
  {J. Chem. Phys.}\ }\textbf {\bibinfo {volume} {123}},\ \bibinfo {pages}
  {034102} (\bibinfo {year} {2005}{\natexlab{b}})}\BibitemShut {NoStop}%
\bibitem [{\citenamefont {Habershon}\ \emph {et~al.}(2007)\citenamefont
  {Habershon}, \citenamefont {Braams},\ and\ \citenamefont
  {Manolopoulos}}]{habershon2007quantum}%
  \BibitemOpen
  \bibfield  {author} {\bibinfo {author} {\bibfnamefont {S.}~\bibnamefont
  {Habershon}}, \bibinfo {author} {\bibfnamefont {B.~J.}\ \bibnamefont
  {Braams}}, \ and\ \bibinfo {author} {\bibfnamefont {D.~E.}\ \bibnamefont
  {Manolopoulos}},\ }\href@noop {} {\bibfield  {journal} {\bibinfo  {journal}
  {J. Chem. Phys.}\ }\textbf {\bibinfo {volume} {127}},\ \bibinfo {pages}
  {174108} (\bibinfo {year} {2007})}\BibitemShut {NoStop}%
\bibitem [{\citenamefont {Markland}\ and\ \citenamefont
  {Manolopoulos}(2008)}]{markland2008efficient}%
  \BibitemOpen
  \bibfield  {author} {\bibinfo {author} {\bibfnamefont {T.~E.}\ \bibnamefont
  {Markland}}\ and\ \bibinfo {author} {\bibfnamefont {D.~E.}\ \bibnamefont
  {Manolopoulos}},\ }\href@noop {} {\bibfield  {journal} {\bibinfo  {journal}
  {J. Chem. Phys.}\ }\textbf {\bibinfo {volume} {129}},\ \bibinfo {pages}
  {024105} (\bibinfo {year} {2008})}\BibitemShut {NoStop}%
\bibitem [{\citenamefont {Richardson}\ and\ \citenamefont
  {Althorpe}(2009)}]{richardson2009ring}%
  \BibitemOpen
  \bibfield  {author} {\bibinfo {author} {\bibfnamefont {J.~O.}\ \bibnamefont
  {Richardson}}\ and\ \bibinfo {author} {\bibfnamefont {S.~C.}\ \bibnamefont
  {Althorpe}},\ }\href@noop {} {\bibfield  {journal} {\bibinfo  {journal} {J.
  Chem. Phys.}\ }\textbf {\bibinfo {volume} {131}},\ \bibinfo {pages} {214106}
  (\bibinfo {year} {2009})}\BibitemShut {NoStop}%
\bibitem [{\citenamefont {Suleimanov}\ \emph {et~al.}(2011)\citenamefont
  {Suleimanov}, \citenamefont {Collepardo-Guevara},\ and\ \citenamefont
  {Manolopoulos}}]{suleimanov2011bimolecular}%
  \BibitemOpen
  \bibfield  {author} {\bibinfo {author} {\bibfnamefont {Y.~V.}\ \bibnamefont
  {Suleimanov}}, \bibinfo {author} {\bibfnamefont {R.}~\bibnamefont
  {Collepardo-Guevara}}, \ and\ \bibinfo {author} {\bibfnamefont {D.~E.}\
  \bibnamefont {Manolopoulos}},\ }\href@noop {} {\bibfield  {journal} {\bibinfo
   {journal} {J. Chem. Phys.}\ }\textbf {\bibinfo {volume} {134}},\ \bibinfo
  {pages} {044131} (\bibinfo {year} {2011})}\BibitemShut {NoStop}%
\bibitem [{\citenamefont {Menzeleev}\ \emph {et~al.}(2011)\citenamefont
  {Menzeleev}, \citenamefont {Ananth},\ and\ \citenamefont {{Miller
  III}}}]{menzeleev2011direct}%
  \BibitemOpen
  \bibfield  {author} {\bibinfo {author} {\bibfnamefont {A.~R.}\ \bibnamefont
  {Menzeleev}}, \bibinfo {author} {\bibfnamefont {N.}~\bibnamefont {Ananth}}, \
  and\ \bibinfo {author} {\bibfnamefont {T.~F.}\ \bibnamefont {{Miller III}}},\
  }\href@noop {} {\bibfield  {journal} {\bibinfo  {journal} {J. Chem. Phys.}\
  }\textbf {\bibinfo {volume} {135}},\ \bibinfo {pages} {074106} (\bibinfo
  {year} {2011})}\BibitemShut {NoStop}%
\bibitem [{\citenamefont {Ananth}(2013)}]{ananth2013mapping}%
  \BibitemOpen
  \bibfield  {author} {\bibinfo {author} {\bibfnamefont {N.}~\bibnamefont
  {Ananth}},\ }\href@noop {} {\bibfield  {journal} {\bibinfo  {journal} {J.
  Chem. Phys.}\ }\textbf {\bibinfo {volume} {139}},\ \bibinfo {pages} {124102}
  (\bibinfo {year} {2013})}\BibitemShut {NoStop}%
\bibitem [{\citenamefont {Shakib}\ and\ \citenamefont
  {Huo}(2017)}]{shakib2017ring}%
  \BibitemOpen
  \bibfield  {author} {\bibinfo {author} {\bibfnamefont {F.~A.}\ \bibnamefont
  {Shakib}}\ and\ \bibinfo {author} {\bibfnamefont {P.}~\bibnamefont {Huo}},\
  }\href@noop {} {\bibfield  {journal} {\bibinfo  {journal} {J. Phys. Chem.
  Lett.}\ }\textbf {\bibinfo {volume} {8}},\ \bibinfo {pages} {3073} (\bibinfo
  {year} {2017})}\BibitemShut {NoStop}%
\bibitem [{\citenamefont {Witt}\ \emph {et~al.}(2009)\citenamefont {Witt},
  \citenamefont {Ivanov}, \citenamefont {Shiga}, \citenamefont {Forbert},\ and\
  \citenamefont {Marx}}]{witt2009applicability}%
  \BibitemOpen
  \bibfield  {author} {\bibinfo {author} {\bibfnamefont {A.}~\bibnamefont
  {Witt}}, \bibinfo {author} {\bibfnamefont {S.~D.}\ \bibnamefont {Ivanov}},
  \bibinfo {author} {\bibfnamefont {M.}~\bibnamefont {Shiga}}, \bibinfo
  {author} {\bibfnamefont {H.}~\bibnamefont {Forbert}}, \ and\ \bibinfo
  {author} {\bibfnamefont {D.}~\bibnamefont {Marx}},\ }\href@noop {} {\bibfield
   {journal} {\bibinfo  {journal} {J. Chem. Phys.}\ }\textbf {\bibinfo {volume}
  {130}},\ \bibinfo {pages} {194510} (\bibinfo {year} {2009})}\BibitemShut
  {NoStop}%
\bibitem [{\citenamefont {Rossi}\ \emph {et~al.}(2014)\citenamefont {Rossi},
  \citenamefont {Ceriotti},\ and\ \citenamefont
  {Manolopoulos}}]{rossi_manolopoulos_TRPDM_2014}%
  \BibitemOpen
  \bibfield  {author} {\bibinfo {author} {\bibfnamefont {M.}~\bibnamefont
  {Rossi}}, \bibinfo {author} {\bibfnamefont {M.}~\bibnamefont {Ceriotti}}, \
  and\ \bibinfo {author} {\bibfnamefont {D.~E.}\ \bibnamefont {Manolopoulos}},\
  }\href@noop {} {\bibfield  {journal} {\bibinfo  {journal} {J. Chem. Phys.}\
  }\textbf {\bibinfo {volume} {140}},\ \bibinfo {pages} {234116} (\bibinfo
  {year} {2014})}\BibitemShut {NoStop}%
\bibitem [{\citenamefont {Miller}(2006)}]{Miller2006}%
  \BibitemOpen
  \bibfield  {author} {\bibinfo {author} {\bibfnamefont {W.~H.}\ \bibnamefont
  {Miller}},\ }\href@noop {} {\bibfield  {journal} {\bibinfo  {journal} {J.
  Chem. Phys.}\ }\textbf {\bibinfo {volume} {125}},\ \bibinfo {pages} {132305}
  (\bibinfo {year} {2006})}\BibitemShut {NoStop}%
\bibitem [{\citenamefont {Heller}(1975)}]{Heller_TdependentSC_1975}%
  \BibitemOpen
  \bibfield  {author} {\bibinfo {author} {\bibfnamefont {E.~J.}\ \bibnamefont
  {Heller}},\ }\href {\doibase 10.1063/1.430620} {\bibfield  {journal}
  {\bibinfo  {journal} {J. Chem. Phys.}\ }\textbf {\bibinfo {volume} {62}},\
  \bibinfo {pages} {1544} (\bibinfo {year} {1975})}\BibitemShut {NoStop}%
\bibitem [{\citenamefont {Kay}(2005)}]{Kay_Atomsandmolecules_2005}%
  \BibitemOpen
  \bibfield  {author} {\bibinfo {author} {\bibfnamefont {K.~G.}\ \bibnamefont
  {Kay}},\ }\href@noop {} {\bibfield  {journal} {\bibinfo  {journal} {Annu.
  Rev. Phys. Chem.}\ }\textbf {\bibinfo {volume} {56}},\ \bibinfo {pages} {255}
  (\bibinfo {year} {2005})}\BibitemShut {NoStop}%
\bibitem [{\citenamefont {Thoss}\ and\ \citenamefont
  {Wang}(2004)}]{Thoss_Wang_SemiclassicalReview_2004}%
  \BibitemOpen
  \bibfield  {author} {\bibinfo {author} {\bibfnamefont {M.}~\bibnamefont
  {Thoss}}\ and\ \bibinfo {author} {\bibfnamefont {H.}~\bibnamefont {Wang}},\
  }\href@noop {} {\bibfield  {journal} {\bibinfo  {journal} {Annu. Rev. Phys.
  Chem.}\ }\textbf {\bibinfo {volume} {55}},\ \bibinfo {pages} {299} (\bibinfo
  {year} {2004})}\BibitemShut {NoStop}%
\bibitem [{\citenamefont {Walton}\ and\ \citenamefont
  {Manolopoulos}(1996)}]{Walton_Manolopoulos_FranckCondon_1996}%
  \BibitemOpen
  \bibfield  {author} {\bibinfo {author} {\bibfnamefont {A.~R.}\ \bibnamefont
  {Walton}}\ and\ \bibinfo {author} {\bibfnamefont {D.~E.}\ \bibnamefont
  {Manolopoulos}},\ }\href@noop {} {\bibfield  {journal} {\bibinfo  {journal}
  {Mol. Phys.}\ }\textbf {\bibinfo {volume} {87}},\ \bibinfo {pages} {961}
  (\bibinfo {year} {1996})}\BibitemShut {NoStop}%
\bibitem [{\citenamefont {Walton}\ and\ \citenamefont
  {Manolopoulos}(1995)}]{Walton_Manolopoulos_FrozenGaussianCO2_1995}%
  \BibitemOpen
  \bibfield  {author} {\bibinfo {author} {\bibfnamefont {A.~R.}\ \bibnamefont
  {Walton}}\ and\ \bibinfo {author} {\bibfnamefont {D.~E.}\ \bibnamefont
  {Manolopoulos}},\ }\href@noop {} {\bibfield  {journal} {\bibinfo  {journal}
  {Chem. Phys. Lett.}\ }\textbf {\bibinfo {volume} {244}},\ \bibinfo {pages}
  {448} (\bibinfo {year} {1995})}\BibitemShut {NoStop}%
\bibitem [{\citenamefont {Pollak}(2007)}]{Pollak_Perturbationseries_2007}%
  \BibitemOpen
  \bibfield  {author} {\bibinfo {author} {\bibfnamefont {E.}~\bibnamefont
  {Pollak}},\ }\enquote {\bibinfo {title} {{The Semiclassical Initial Value
  Series Representation of the Quantum Propagator}},}\ in\ \href {\doibase
  10.1007/978-3-540-34460-5_11} {\emph {\bibinfo {booktitle} {{Quantum Dynamics
  of Complex Molecular Systems}}}},\ \bibinfo {editor} {edited by\ \bibinfo
  {editor} {\bibfnamefont {D.~A.}\ \bibnamefont {Micha}}\ and\ \bibinfo
  {editor} {\bibfnamefont {I.}~\bibnamefont {Burghardt}}}\ (\bibinfo
  {publisher} {Springer Berlin Heidelberg},\ \bibinfo {address} {Berlin,
  Heidelberg},\ \bibinfo {year} {2007})\ pp.\ \bibinfo {pages}
  {259--271}\BibitemShut {NoStop}%
\bibitem [{\citenamefont {Bonella}\ \emph {et~al.}(2005)\citenamefont
  {Bonella}, \citenamefont {Montemayor},\ and\ \citenamefont
  {Coker}}]{Bonella_Coker_Linearizedpathintegral_2005}%
  \BibitemOpen
  \bibfield  {author} {\bibinfo {author} {\bibfnamefont {S.}~\bibnamefont
  {Bonella}}, \bibinfo {author} {\bibfnamefont {D.}~\bibnamefont {Montemayor}},
  \ and\ \bibinfo {author} {\bibfnamefont {D.~F.}\ \bibnamefont {Coker}},\
  }\href@noop {} {\bibfield  {journal} {\bibinfo  {journal} {Proc. Natl. Acad.
  Sci.}\ }\textbf {\bibinfo {volume} {102}},\ \bibinfo {pages} {6715} (\bibinfo
  {year} {2005})}\BibitemShut {NoStop}%
\bibitem [{\citenamefont {Huo}\ and\ \citenamefont
  {Coker}(2012)}]{Huo_Coker_Semiclassicalnonadiabatic_2012}%
  \BibitemOpen
  \bibfield  {author} {\bibinfo {author} {\bibfnamefont {P.}~\bibnamefont
  {Huo}}\ and\ \bibinfo {author} {\bibfnamefont {D.~F.}\ \bibnamefont
  {Coker}},\ }\href@noop {} {\bibfield  {journal} {\bibinfo  {journal} {Mol.
  Phys.}\ }\textbf {\bibinfo {volume} {110}},\ \bibinfo {pages} {1035}
  (\bibinfo {year} {2012})}\BibitemShut {NoStop}%
\bibitem [{\citenamefont
  {Miller}(2001{\natexlab{a}})}]{Miller_Addingquantumtoclassical_2001}%
  \BibitemOpen
  \bibfield  {author} {\bibinfo {author} {\bibfnamefont {W.~H.}\ \bibnamefont
  {Miller}},\ }\href@noop {} {\bibfield  {journal} {\bibinfo  {journal} {J.
  Phys. Chem. A}\ }\textbf {\bibinfo {volume} {105}},\ \bibinfo {pages} {2942}
  (\bibinfo {year} {2001}{\natexlab{a}})}\BibitemShut {NoStop}%
\bibitem [{\citenamefont {Miller}(2005)}]{Miller_PNAScomplexsystems_2005}%
  \BibitemOpen
  \bibfield  {author} {\bibinfo {author} {\bibfnamefont {W.~H.}\ \bibnamefont
  {Miller}},\ }\href {\doibase 10.1073/pnas.0408043102} {\bibfield  {journal}
  {\bibinfo  {journal} {Proc. Natl. Acad. Sci. USA}\ }\textbf {\bibinfo
  {volume} {102}},\ \bibinfo {pages} {6660} (\bibinfo {year}
  {2005})}\BibitemShut {NoStop}%
\bibitem [{\citenamefont {Kay}(1994{\natexlab{a}})}]{Kay_Multidim_1994}%
  \BibitemOpen
  \bibfield  {author} {\bibinfo {author} {\bibfnamefont {K.~G.}\ \bibnamefont
  {Kay}},\ }\href@noop {} {\bibfield  {journal} {\bibinfo  {journal} {J. Chem.
  Phys.}\ }\textbf {\bibinfo {volume} {101}},\ \bibinfo {pages} {2250}
  (\bibinfo {year} {1994}{\natexlab{a}})}\BibitemShut {NoStop}%
\bibitem [{\citenamefont {Kay}(1994{\natexlab{b}})}]{Kay_Numerical_1994}%
  \BibitemOpen
  \bibfield  {author} {\bibinfo {author} {\bibfnamefont {K.~G.}\ \bibnamefont
  {Kay}},\ }\href@noop {} {\bibfield  {journal} {\bibinfo  {journal} {J. Chem.
  Phys.}\ }\textbf {\bibinfo {volume} {100}},\ \bibinfo {pages} {4432}
  (\bibinfo {year} {1994}{\natexlab{b}})}\BibitemShut {NoStop}%
\bibitem [{\citenamefont
  {Heller}(1981{\natexlab{a}})}]{Heller_FrozenGaussian_1981}%
  \BibitemOpen
  \bibfield  {author} {\bibinfo {author} {\bibfnamefont {E.~J.}\ \bibnamefont
  {Heller}},\ }\href {\doibase 10.1063/1.442382} {\bibfield  {journal}
  {\bibinfo  {journal} {J. Chem. Phys.}\ }\textbf {\bibinfo {volume} {75}},\
  \bibinfo {pages} {2923} (\bibinfo {year} {1981}{\natexlab{a}})}\BibitemShut
  {NoStop}%
\bibitem [{\citenamefont {Wang}\ \emph {et~al.}(1998)\citenamefont {Wang},
  \citenamefont {Sun},\ and\ \citenamefont
  {Miller}}]{Wang_Miller_Chemicalreactions_1998}%
  \BibitemOpen
  \bibfield  {author} {\bibinfo {author} {\bibfnamefont {H.}~\bibnamefont
  {Wang}}, \bibinfo {author} {\bibfnamefont {X.}~\bibnamefont {Sun}}, \ and\
  \bibinfo {author} {\bibfnamefont {W.~H.}\ \bibnamefont {Miller}},\
  }\href@noop {} {\bibfield  {journal} {\bibinfo  {journal} {J. Chem. Phys.}\
  }\textbf {\bibinfo {volume} {108}},\ \bibinfo {pages} {9726} (\bibinfo {year}
  {1998})}\BibitemShut {NoStop}%
\bibitem [{\citenamefont {Yamamoto}\ \emph {et~al.}(2002)\citenamefont
  {Yamamoto}, \citenamefont {Wang},\ and\ \citenamefont
  {Miller}}]{Yamamoto_Miller_Fluxcorrelation_2002}%
  \BibitemOpen
  \bibfield  {author} {\bibinfo {author} {\bibfnamefont {T.}~\bibnamefont
  {Yamamoto}}, \bibinfo {author} {\bibfnamefont {H.}~\bibnamefont {Wang}}, \
  and\ \bibinfo {author} {\bibfnamefont {W.~H.}\ \bibnamefont {Miller}},\
  }\href@noop {} {\bibfield  {journal} {\bibinfo  {journal} {J. Chem. Phys.}\
  }\textbf {\bibinfo {volume} {116}},\ \bibinfo {pages} {7335} (\bibinfo {year}
  {2002})}\BibitemShut {NoStop}%
\bibitem [{\citenamefont {Conte}\ and\ \citenamefont
  {Pollak}(2012)}]{Conte_Pollak_ContinuumLimit_2012}%
  \BibitemOpen
  \bibfield  {author} {\bibinfo {author} {\bibfnamefont {R.}~\bibnamefont
  {Conte}}\ and\ \bibinfo {author} {\bibfnamefont {E.}~\bibnamefont {Pollak}},\
  }\href@noop {} {\bibfield  {journal} {\bibinfo  {journal} {J. Chem. Phys.}\
  }\textbf {\bibinfo {volume} {136}},\ \bibinfo {pages} {094101} (\bibinfo
  {year} {2012})}\BibitemShut {NoStop}%
\bibitem [{\citenamefont {Conte}\ and\ \citenamefont
  {Pollak}(2010)}]{Conte_Pollak_ThawedGaussian_2010}%
  \BibitemOpen
  \bibfield  {author} {\bibinfo {author} {\bibfnamefont {R.}~\bibnamefont
  {Conte}}\ and\ \bibinfo {author} {\bibfnamefont {E.}~\bibnamefont {Pollak}},\
  }\href@noop {} {\bibfield  {journal} {\bibinfo  {journal} {Phys. Rev. E}\
  }\textbf {\bibinfo {volume} {81}},\ \bibinfo {pages} {036704} (\bibinfo
  {year} {2010})}\BibitemShut {NoStop}%
\bibitem [{\citenamefont {Berry}\ and\ \citenamefont
  {Mount}(1972)}]{Berry_Mount_Semiclassical_1972}%
  \BibitemOpen
  \bibfield  {author} {\bibinfo {author} {\bibfnamefont {M.~V.}\ \bibnamefont
  {Berry}}\ and\ \bibinfo {author} {\bibfnamefont {K.}~\bibnamefont {Mount}},\
  }\href@noop {} {\bibfield  {journal} {\bibinfo  {journal} {Rep. on Prog.
  Phys.}\ }\textbf {\bibinfo {volume} {35}},\ \bibinfo {pages} {315} (\bibinfo
  {year} {1972})}\BibitemShut {NoStop}%
\bibitem [{\citenamefont {Wang}\ \emph {et~al.}(2000)\citenamefont {Wang},
  \citenamefont {Thoss},\ and\ \citenamefont {Miller}}]{Wang2000}%
  \BibitemOpen
  \bibfield  {author} {\bibinfo {author} {\bibfnamefont {H.}~\bibnamefont
  {Wang}}, \bibinfo {author} {\bibfnamefont {M.}~\bibnamefont {Thoss}}, \ and\
  \bibinfo {author} {\bibfnamefont {W.~H.}\ \bibnamefont {Miller}},\ }\href
  {\doibase 10.1063/1.480560} {\bibfield  {journal} {\bibinfo  {journal} {J.
  Chem. Phys.}\ }\textbf {\bibinfo {volume} {112}},\ \bibinfo {pages} {47}
  (\bibinfo {year} {2000})}\BibitemShut {NoStop}%
\bibitem [{\citenamefont {Wang}\ \emph {et~al.}(2001)\citenamefont {Wang},
  \citenamefont {Thoss}, \citenamefont {Sorge}, \citenamefont {Gelabert},
  \citenamefont {Gim{\'e}nez},\ and\ \citenamefont {Miller}}]{Wang2001}%
  \BibitemOpen
  \bibfield  {author} {\bibinfo {author} {\bibfnamefont {H.}~\bibnamefont
  {Wang}}, \bibinfo {author} {\bibfnamefont {M.}~\bibnamefont {Thoss}},
  \bibinfo {author} {\bibfnamefont {K.~L.}\ \bibnamefont {Sorge}}, \bibinfo
  {author} {\bibfnamefont {R.}~\bibnamefont {Gelabert}}, \bibinfo {author}
  {\bibfnamefont {X.}~\bibnamefont {Gim{\'e}nez}}, \ and\ \bibinfo {author}
  {\bibfnamefont {W.~H.}\ \bibnamefont {Miller}},\ }\href {\doibase
  10.1063/1.1337802} {\bibfield  {journal} {\bibinfo  {journal} {J. Chem.
  Phys.}\ }\textbf {\bibinfo {volume} {114}},\ \bibinfo {pages} {2562}
  (\bibinfo {year} {2001})}\BibitemShut {NoStop}%
\bibitem [{\citenamefont {Zhuang}\ \emph {et~al.}(2012)\citenamefont {Zhuang},
  \citenamefont {Siebert}, \citenamefont {Hase}, \citenamefont {Kay},\ and\
  \citenamefont {Ceotto}}]{Zhuang_Ceotto_Hessianapprox_2012}%
  \BibitemOpen
  \bibfield  {author} {\bibinfo {author} {\bibfnamefont {Y.}~\bibnamefont
  {Zhuang}}, \bibinfo {author} {\bibfnamefont {M.~R.}\ \bibnamefont {Siebert}},
  \bibinfo {author} {\bibfnamefont {W.~L.}\ \bibnamefont {Hase}}, \bibinfo
  {author} {\bibfnamefont {K.~G.}\ \bibnamefont {Kay}}, \ and\ \bibinfo
  {author} {\bibfnamefont {M.}~\bibnamefont {Ceotto}},\ }\href@noop {}
  {\bibfield  {journal} {\bibinfo  {journal} {J. Chem. Theory Comput.}\
  }\textbf {\bibinfo {volume} {9}},\ \bibinfo {pages} {54} (\bibinfo {year}
  {2012})}\BibitemShut {NoStop}%
\bibitem [{\citenamefont {Ceotto}\ \emph {et~al.}(2017)\citenamefont {Ceotto},
  \citenamefont {{Di Liberto}},\ and\ \citenamefont
  {Conte}}]{ceotto_conte_DCSCIVR_2017}%
  \BibitemOpen
  \bibfield  {author} {\bibinfo {author} {\bibfnamefont {M.}~\bibnamefont
  {Ceotto}}, \bibinfo {author} {\bibfnamefont {G.}~\bibnamefont {{Di
  Liberto}}}, \ and\ \bibinfo {author} {\bibfnamefont {R.}~\bibnamefont
  {Conte}},\ }\href@noop {} {\bibfield  {journal} {\bibinfo  {journal} {Phys.
  Rev. Lett.}\ }\textbf {\bibinfo {volume} {119}},\ \bibinfo {pages} {010401}
  (\bibinfo {year} {2017})}\BibitemShut {NoStop}%
\bibitem [{\citenamefont {Gabas}\ \emph {et~al.}(2017)\citenamefont {Gabas},
  \citenamefont {Conte},\ and\ \citenamefont
  {Ceotto}}]{Gabas_Ceotto_Glycine_2017}%
  \BibitemOpen
  \bibfield  {author} {\bibinfo {author} {\bibfnamefont {F.}~\bibnamefont
  {Gabas}}, \bibinfo {author} {\bibfnamefont {R.}~\bibnamefont {Conte}}, \ and\
  \bibinfo {author} {\bibfnamefont {M.}~\bibnamefont {Ceotto}},\ }\href@noop {}
  {\bibfield  {journal} {\bibinfo  {journal} {J. Chem. Theory Comput.}\
  }\textbf {\bibinfo {volume} {13}},\ \bibinfo {pages} {2378} (\bibinfo {year}
  {2017})}\BibitemShut {NoStop}%
\bibitem [{\citenamefont {Monteferrante}\ \emph {et~al.}(2013)\citenamefont
  {Monteferrante}, \citenamefont {Bonella},\ and\ \citenamefont
  {Ciccotti}}]{Monteferrante_Ciccotti_Liquidneon_2013}%
  \BibitemOpen
  \bibfield  {author} {\bibinfo {author} {\bibfnamefont {M.}~\bibnamefont
  {Monteferrante}}, \bibinfo {author} {\bibfnamefont {S.}~\bibnamefont
  {Bonella}}, \ and\ \bibinfo {author} {\bibfnamefont {G.}~\bibnamefont
  {Ciccotti}},\ }\href@noop {} {\bibfield  {journal} {\bibinfo  {journal} {J.
  Chem. Phys.}\ }\textbf {\bibinfo {volume} {138}},\ \bibinfo {pages} {054118}
  (\bibinfo {year} {2013})}\BibitemShut {NoStop}%
\bibitem [{\citenamefont {Bonella}\ \emph {et~al.}(2010)\citenamefont
  {Bonella}, \citenamefont {Ciccotti},\ and\ \citenamefont
  {Kapral}}]{Bonella_Kapral_quantum-classical_2010}%
  \BibitemOpen
  \bibfield  {author} {\bibinfo {author} {\bibfnamefont {S.}~\bibnamefont
  {Bonella}}, \bibinfo {author} {\bibfnamefont {G.}~\bibnamefont {Ciccotti}}, \
  and\ \bibinfo {author} {\bibfnamefont {R.}~\bibnamefont {Kapral}},\
  }\href@noop {} {\bibfield  {journal} {\bibinfo  {journal} {Chem. Phys.
  Lett.}\ }\textbf {\bibinfo {volume} {484}},\ \bibinfo {pages} {399} (\bibinfo
  {year} {2010})}\BibitemShut {NoStop}%
\bibitem [{\citenamefont {Pollak}\ and\ \citenamefont
  {Martin-Fierro}(2007)}]{Pollak_MFierro_FBIVR_2007}%
  \BibitemOpen
  \bibfield  {author} {\bibinfo {author} {\bibfnamefont {E.}~\bibnamefont
  {Pollak}}\ and\ \bibinfo {author} {\bibfnamefont {E.}~\bibnamefont
  {Martin-Fierro}},\ }\href@noop {} {\bibfield  {journal} {\bibinfo  {journal}
  {J. Chem. Phys.}\ }\textbf {\bibinfo {volume} {126}},\ \bibinfo {pages}
  {164107} (\bibinfo {year} {2007})}\BibitemShut {NoStop}%
\bibitem [{\citenamefont {Wehrle}\ \emph {et~al.}(2015)\citenamefont {Wehrle},
  \citenamefont {Oberli},\ and\ \citenamefont
  {{Van{\'i}\v{c}ek}}}]{Wehrle_Vanicek_NH3_2015}%
  \BibitemOpen
  \bibfield  {author} {\bibinfo {author} {\bibfnamefont {M.}~\bibnamefont
  {Wehrle}}, \bibinfo {author} {\bibfnamefont {S.}~\bibnamefont {Oberli}}, \
  and\ \bibinfo {author} {\bibfnamefont {J.}~\bibnamefont
  {{Van{\'i}\v{c}ek}}},\ }\href {\doibase 10.1021/acs.jpca.5b03907} {\bibfield
  {journal} {\bibinfo  {journal} {J. Phys. Chem. A}\ }\textbf {\bibinfo
  {volume} {119}},\ \bibinfo {pages} {5685} (\bibinfo {year}
  {2015})}\BibitemShut {NoStop}%
\bibitem [{\citenamefont {Wehrle}\ \emph {et~al.}(2014)\citenamefont {Wehrle},
  \citenamefont {Sulc},\ and\ \citenamefont
  {Vanicek}}]{Wehrle_Vanicek_Oligothiophenes_2014}%
  \BibitemOpen
  \bibfield  {author} {\bibinfo {author} {\bibfnamefont {M.}~\bibnamefont
  {Wehrle}}, \bibinfo {author} {\bibfnamefont {M.}~\bibnamefont {Sulc}}, \ and\
  \bibinfo {author} {\bibfnamefont {J.}~\bibnamefont {Vanicek}},\ }\href
  {\doibase 10.1063/1.4884718} {\bibfield  {journal} {\bibinfo  {journal} {J.
  Chem. Phys.}\ }\textbf {\bibinfo {volume} {140}},\ \bibinfo {pages} {244114}
  (\bibinfo {year} {2014})}\BibitemShut {NoStop}%
\bibitem [{\citenamefont {Ceotto}\ \emph
  {et~al.}(2009{\natexlab{a}})\citenamefont {Ceotto}, \citenamefont {Atahan},
  \citenamefont {Tantardini},\ and\ \citenamefont
  {Aspuru-Guzik}}]{Ceotto_AspuruGuzik_Multiplecoherent_2009}%
  \BibitemOpen
  \bibfield  {author} {\bibinfo {author} {\bibfnamefont {M.}~\bibnamefont
  {Ceotto}}, \bibinfo {author} {\bibfnamefont {S.}~\bibnamefont {Atahan}},
  \bibinfo {author} {\bibfnamefont {G.~F.}\ \bibnamefont {Tantardini}}, \ and\
  \bibinfo {author} {\bibfnamefont {A.}~\bibnamefont {Aspuru-Guzik}},\ }\href
  {\doibase 10.1063/1.3155062} {\bibfield  {journal} {\bibinfo  {journal} {J.
  Chem. Phys.}\ }\textbf {\bibinfo {volume} {130}},\ \bibinfo {pages} {234113}
  (\bibinfo {year} {2009}{\natexlab{a}})}\BibitemShut {NoStop}%
\bibitem [{\citenamefont {Ceotto}\ \emph
  {et~al.}(2009{\natexlab{b}})\citenamefont {Ceotto}, \citenamefont {Atahan},
  \citenamefont {Shim}, \citenamefont {Tantardini},\ and\ \citenamefont
  {Aspuru-Guzik}}]{Ceotto_AspuruGuzik_PCCPFirstprinciples_2009}%
  \BibitemOpen
  \bibfield  {author} {\bibinfo {author} {\bibfnamefont {M.}~\bibnamefont
  {Ceotto}}, \bibinfo {author} {\bibfnamefont {S.}~\bibnamefont {Atahan}},
  \bibinfo {author} {\bibfnamefont {S.}~\bibnamefont {Shim}}, \bibinfo {author}
  {\bibfnamefont {G.~F.}\ \bibnamefont {Tantardini}}, \ and\ \bibinfo {author}
  {\bibfnamefont {A.}~\bibnamefont {Aspuru-Guzik}},\ }\href {\doibase
  10.1039/B820785B} {\bibfield  {journal} {\bibinfo  {journal} {Phys. Chem.
  Chem. Phys.}\ }\textbf {\bibinfo {volume} {11}},\ \bibinfo {pages} {3861}
  (\bibinfo {year} {2009}{\natexlab{b}})}\BibitemShut {NoStop}%
\bibitem [{\citenamefont {Ceotto}\ \emph {et~al.}(2010)\citenamefont {Ceotto},
  \citenamefont {{Dell` Angelo}},\ and\ \citenamefont
  {Tantardini}}]{Ceotto_Tantardini_Copper100_2010}%
  \BibitemOpen
  \bibfield  {author} {\bibinfo {author} {\bibfnamefont {M.}~\bibnamefont
  {Ceotto}}, \bibinfo {author} {\bibfnamefont {D.}~\bibnamefont {{Dell`
  Angelo}}}, \ and\ \bibinfo {author} {\bibfnamefont {G.~F.}\ \bibnamefont
  {Tantardini}},\ }\href@noop {} {\bibfield  {journal} {\bibinfo  {journal} {J.
  Chem. Phys.}\ }\textbf {\bibinfo {volume} {133}},\ \bibinfo {pages} {054701}
  (\bibinfo {year} {2010})}\BibitemShut {NoStop}%
\bibitem [{\citenamefont {Ceotto}\ \emph
  {et~al.}(2011{\natexlab{a}})\citenamefont {Ceotto}, \citenamefont
  {Tantardini},\ and\ \citenamefont
  {Aspuru-Guzik}}]{Ceotto_AspuruGuzik_Curseofdimensionality_2011}%
  \BibitemOpen
  \bibfield  {author} {\bibinfo {author} {\bibfnamefont {M.}~\bibnamefont
  {Ceotto}}, \bibinfo {author} {\bibfnamefont {G.~F.}\ \bibnamefont
  {Tantardini}}, \ and\ \bibinfo {author} {\bibfnamefont {A.}~\bibnamefont
  {Aspuru-Guzik}},\ }\href {\doibase 10.1063/1.3664731} {\bibfield  {journal}
  {\bibinfo  {journal} {J. Chem. Phys.}\ }\textbf {\bibinfo {volume} {135}},\
  \bibinfo {pages} {214108} (\bibinfo {year} {2011}{\natexlab{a}})}\BibitemShut
  {NoStop}%
\bibitem [{\citenamefont {Ceotto}\ \emph
  {et~al.}(2011{\natexlab{b}})\citenamefont {Ceotto}, \citenamefont {Valleau},
  \citenamefont {Tantardini},\ and\ \citenamefont
  {Aspuru-Guzik}}]{Ceotto_AspuruGuzik_Firstprinciples_2011}%
  \BibitemOpen
  \bibfield  {author} {\bibinfo {author} {\bibfnamefont {M.}~\bibnamefont
  {Ceotto}}, \bibinfo {author} {\bibfnamefont {S.}~\bibnamefont {Valleau}},
  \bibinfo {author} {\bibfnamefont {G.~F.}\ \bibnamefont {Tantardini}}, \ and\
  \bibinfo {author} {\bibfnamefont {A.}~\bibnamefont {Aspuru-Guzik}},\ }\href
  {\doibase 10.1063/1.3599469} {\bibfield  {journal} {\bibinfo  {journal} {J.
  Chem. Phys.}\ }\textbf {\bibinfo {volume} {134}},\ \bibinfo {pages} {234103}
  (\bibinfo {year} {2011}{\natexlab{b}})}\BibitemShut {NoStop}%
\bibitem [{\citenamefont {Tamascelli}\ \emph {et~al.}(2014)\citenamefont
  {Tamascelli}, \citenamefont {Dambrosio}, \citenamefont {Conte},\ and\
  \citenamefont {Ceotto}}]{Tamascelli_Ceotto_GPU_2014}%
  \BibitemOpen
  \bibfield  {author} {\bibinfo {author} {\bibfnamefont {D.}~\bibnamefont
  {Tamascelli}}, \bibinfo {author} {\bibfnamefont {F.~S.}\ \bibnamefont
  {Dambrosio}}, \bibinfo {author} {\bibfnamefont {R.}~\bibnamefont {Conte}}, \
  and\ \bibinfo {author} {\bibfnamefont {M.}~\bibnamefont {Ceotto}},\
  }\href@noop {} {\bibfield  {journal} {\bibinfo  {journal} {J. Chem. Phys.}\
  }\textbf {\bibinfo {volume} {140}},\ \bibinfo {pages} {174109} (\bibinfo
  {year} {2014})}\BibitemShut {NoStop}%
\bibitem [{\citenamefont {Conte}\ \emph
  {et~al.}(2013{\natexlab{b}})\citenamefont {Conte}, \citenamefont
  {Aspuru-Guzik},\ and\ \citenamefont {Ceotto}}]{Conte_Ceotto_NH3_2013}%
  \BibitemOpen
  \bibfield  {author} {\bibinfo {author} {\bibfnamefont {R.}~\bibnamefont
  {Conte}}, \bibinfo {author} {\bibfnamefont {A.}~\bibnamefont {Aspuru-Guzik}},
  \ and\ \bibinfo {author} {\bibfnamefont {M.}~\bibnamefont {Ceotto}},\ }\href
  {\doibase 10.1021/jz401603f} {\bibfield  {journal} {\bibinfo  {journal} {J.
  Phys. Chem. Lett.}\ }\textbf {\bibinfo {volume} {4}},\ \bibinfo {pages}
  {3407} (\bibinfo {year} {2013}{\natexlab{b}})}\BibitemShut {NoStop}%
\bibitem [{\citenamefont {Ceotto}\ \emph {et~al.}(2013)\citenamefont {Ceotto},
  \citenamefont {Zhuang},\ and\ \citenamefont
  {Hase}}]{Ceotto_Hase_AcceleratedSC_2013}%
  \BibitemOpen
  \bibfield  {author} {\bibinfo {author} {\bibfnamefont {M.}~\bibnamefont
  {Ceotto}}, \bibinfo {author} {\bibfnamefont {Y.}~\bibnamefont {Zhuang}}, \
  and\ \bibinfo {author} {\bibfnamefont {W.~L.}\ \bibnamefont {Hase}},\
  }\href@noop {} {\bibfield  {journal} {\bibinfo  {journal} {J. Chem. Phys.}\
  }\textbf {\bibinfo {volume} {138}},\ \bibinfo {pages} {054116} (\bibinfo
  {year} {2013})}\BibitemShut {NoStop}%
\bibitem [{\citenamefont {Tatchen}\ and\ \citenamefont
  {Pollak}(2009)}]{Tatchen_Pollak_Onthefly_2009}%
  \BibitemOpen
  \bibfield  {author} {\bibinfo {author} {\bibfnamefont {J.}~\bibnamefont
  {Tatchen}}\ and\ \bibinfo {author} {\bibfnamefont {E.}~\bibnamefont
  {Pollak}},\ }\href {\doibase 10.1063/1.3074100} {\bibfield  {journal}
  {\bibinfo  {journal} {J. Chem. Phys.}\ }\textbf {\bibinfo {volume} {130}},\
  \bibinfo {pages} {041103} (\bibinfo {year} {2009})}\BibitemShut {NoStop}%
\bibitem [{\citenamefont {Wong}\ \emph {et~al.}(2011)\citenamefont {Wong},
  \citenamefont {Benoit}, \citenamefont {Lewerenz}, \citenamefont {Brown},\
  and\ \citenamefont {Roy}}]{Wong_Roy_Formaldehyde_2011}%
  \BibitemOpen
  \bibfield  {author} {\bibinfo {author} {\bibfnamefont {S.~Y.~Y.}\
  \bibnamefont {Wong}}, \bibinfo {author} {\bibfnamefont {D.~M.}\ \bibnamefont
  {Benoit}}, \bibinfo {author} {\bibfnamefont {M.}~\bibnamefont {Lewerenz}},
  \bibinfo {author} {\bibfnamefont {A.}~\bibnamefont {Brown}}, \ and\ \bibinfo
  {author} {\bibfnamefont {P.-N.}\ \bibnamefont {Roy}},\ }\href {\doibase
  10.1063/1.3553179} {\bibfield  {journal} {\bibinfo  {journal} {J. Chem.
  Phys.}\ }\textbf {\bibinfo {volume} {134}},\ \bibinfo {pages} {094110}
  (\bibinfo {year} {2011})}\BibitemShut {NoStop}%
\bibitem [{\citenamefont {Thoss}\ \emph {et~al.}(2001)\citenamefont {Thoss},
  \citenamefont {Wang},\ and\ \citenamefont {Miller}}]{thoss2001self}%
  \BibitemOpen
  \bibfield  {author} {\bibinfo {author} {\bibfnamefont {M.}~\bibnamefont
  {Thoss}}, \bibinfo {author} {\bibfnamefont {H.}~\bibnamefont {Wang}}, \ and\
  \bibinfo {author} {\bibfnamefont {W.~H.}\ \bibnamefont {Miller}},\
  }\href@noop {} {\bibfield  {journal} {\bibinfo  {journal} {J. Chem. Phys.}\
  }\textbf {\bibinfo {volume} {115}},\ \bibinfo {pages} {2991} (\bibinfo {year}
  {2001})}\BibitemShut {NoStop}%
\bibitem [{\citenamefont {Buchholz}\ \emph {et~al.}(2017)\citenamefont
  {Buchholz}, \citenamefont {Grossmann},\ and\ \citenamefont
  {Ceotto}}]{Ceotto_Buchholz_MixedSC_2017}%
  \BibitemOpen
  \bibfield  {author} {\bibinfo {author} {\bibfnamefont {M.}~\bibnamefont
  {Buchholz}}, \bibinfo {author} {\bibfnamefont {F.}~\bibnamefont {Grossmann}},
  \ and\ \bibinfo {author} {\bibfnamefont {M.}~\bibnamefont {Ceotto}},\
  }\href@noop {} {\bibfield  {journal} {\bibinfo  {journal} {J. Chem. Phys.}\
  }\textbf {\bibinfo {volume} {147}},\ \bibinfo {pages} {164110} (\bibinfo
  {year} {2017})}\BibitemShut {NoStop}%
\bibitem [{\citenamefont {Miller}(1998)}]{miller1998spiers}%
  \BibitemOpen
  \bibfield  {author} {\bibinfo {author} {\bibfnamefont {W.~H.}\ \bibnamefont
  {Miller}},\ }\href@noop {} {\bibfield  {journal} {\bibinfo  {journal}
  {Faraday Discussions}\ }\textbf {\bibinfo {volume} {110}},\ \bibinfo {pages}
  {1} (\bibinfo {year} {1998})}\BibitemShut {NoStop}%
\bibitem [{\citenamefont
  {Miller}(2001{\natexlab{b}})}]{miller2001semiclassical}%
  \BibitemOpen
  \bibfield  {author} {\bibinfo {author} {\bibfnamefont {W.~H.}\ \bibnamefont
  {Miller}},\ }\href@noop {} {\bibfield  {journal} {\bibinfo  {journal} {J.
  Phys. Chem. A}\ }\textbf {\bibinfo {volume} {105}},\ \bibinfo {pages} {2942}
  (\bibinfo {year} {2001}{\natexlab{b}})}\BibitemShut {NoStop}%
\bibitem [{\citenamefont {Heller}(1991)}]{Heller_Cellulardynamics_1991}%
  \BibitemOpen
  \bibfield  {author} {\bibinfo {author} {\bibfnamefont {E.~J.}\ \bibnamefont
  {Heller}},\ }\href {\doibase 10.1063/1.459848} {\bibfield  {journal}
  {\bibinfo  {journal} {J. Chem. Phys.}\ }\textbf {\bibinfo {volume} {94}},\
  \bibinfo {pages} {2723} (\bibinfo {year} {1991})}\BibitemShut {NoStop}%
\bibitem [{\citenamefont
  {Heller}(1981{\natexlab{b}})}]{Heller_SCspectroscopy_1981}%
  \BibitemOpen
  \bibfield  {author} {\bibinfo {author} {\bibfnamefont {E.~J.}\ \bibnamefont
  {Heller}},\ }\href@noop {} {\bibfield  {journal} {\bibinfo  {journal} {Acc.
  Chem. Res.}\ }\textbf {\bibinfo {volume} {14}},\ \bibinfo {pages} {368}
  (\bibinfo {year} {1981}{\natexlab{b}})}\BibitemShut {NoStop}%
\bibitem [{\citenamefont {Herman}(1994)}]{Herman1994}%
  \BibitemOpen
  \bibfield  {author} {\bibinfo {author} {\bibfnamefont {M.~F.}\ \bibnamefont
  {Herman}},\ }\href@noop {} {\bibfield  {journal} {\bibinfo  {journal} {Annu.
  Rev. Phys. Chem.}\ }\textbf {\bibinfo {volume} {45}},\ \bibinfo {pages} {83}
  (\bibinfo {year} {1994})}\BibitemShut {NoStop}%
\bibitem [{\citenamefont {Herman}\ and\ \citenamefont
  {Kluk}(1984)}]{Herman_Kluk_SCnonspreading_1984}%
  \BibitemOpen
  \bibfield  {author} {\bibinfo {author} {\bibfnamefont {M.~F.}\ \bibnamefont
  {Herman}}\ and\ \bibinfo {author} {\bibfnamefont {E.}~\bibnamefont {Kluk}},\
  }\href {\doibase 10.1016/0301-0104(84)80039-7} {\bibfield  {journal}
  {\bibinfo  {journal} {Chem. Phys.}\ }\textbf {\bibinfo {volume} {91}},\
  \bibinfo {pages} {27} (\bibinfo {year} {1984})}\BibitemShut {NoStop}%
\bibitem [{\citenamefont
  {Kay}(1994{\natexlab{c}})}]{Kay_Integralexpression_1994}%
  \BibitemOpen
  \bibfield  {author} {\bibinfo {author} {\bibfnamefont {K.~G.}\ \bibnamefont
  {Kay}},\ }\href@noop {} {\bibfield  {journal} {\bibinfo  {journal} {J. Chem.
  Phys.}\ }\textbf {\bibinfo {volume} {100}},\ \bibinfo {pages} {4377}
  (\bibinfo {year} {1994}{\natexlab{c}})}\BibitemShut {NoStop}%
\bibitem [{\citenamefont {Kay}(2006)}]{Kay_SCcorrections_2006}%
  \BibitemOpen
  \bibfield  {author} {\bibinfo {author} {\bibfnamefont {K.~G.}\ \bibnamefont
  {Kay}},\ }\href {\doibase 10.1016/j.chemphys.2005.06.019} {\bibfield
  {journal} {\bibinfo  {journal} {Chem. Phys.}\ }\textbf {\bibinfo {volume}
  {322}},\ \bibinfo {pages} {3} (\bibinfo {year} {2006})}\BibitemShut {NoStop}%
\bibitem [{\citenamefont {Martin-Fierro}\ and\ \citenamefont
  {Pollak}(2006)}]{MFierro_Pollak_FBIVR_2006}%
  \BibitemOpen
  \bibfield  {author} {\bibinfo {author} {\bibfnamefont {E.}~\bibnamefont
  {Martin-Fierro}}\ and\ \bibinfo {author} {\bibfnamefont {E.}~\bibnamefont
  {Pollak}},\ }\href@noop {} {\bibfield  {journal} {\bibinfo  {journal} {J.
  Chem. Phys.}\ }\textbf {\bibinfo {volume} {125}},\ \bibinfo {pages} {164104}
  (\bibinfo {year} {2006})}\BibitemShut {NoStop}%
\bibitem [{\citenamefont {Grossmann}(1995)}]{Grossmann1995}%
  \BibitemOpen
  \bibfield  {author} {\bibinfo {author} {\bibfnamefont {F.}~\bibnamefont
  {Grossmann}},\ }\href@noop {} {\bibfield  {journal} {\bibinfo  {journal} {J.
  Chem. Phys.}\ }\textbf {\bibinfo {volume} {103}},\ \bibinfo {pages} {3696}
  (\bibinfo {year} {1995})}\BibitemShut {NoStop}%
\bibitem [{\citenamefont {Grossmann}(1999)}]{Grossmann_HierarchySC_1999}%
  \BibitemOpen
  \bibfield  {author} {\bibinfo {author} {\bibfnamefont {F.}~\bibnamefont
  {Grossmann}},\ }\href@noop {} {\bibfield  {journal} {\bibinfo  {journal}
  {Comments At. Mol. Phys.}\ }\textbf {\bibinfo {volume} {34}},\ \bibinfo
  {pages} {141} (\bibinfo {year} {1999})}\BibitemShut {NoStop}%
\bibitem [{\citenamefont {Sun}\ and\ \citenamefont {Miller}(1997)}]{Sun1997}%
  \BibitemOpen
  \bibfield  {author} {\bibinfo {author} {\bibfnamefont {X.}~\bibnamefont
  {Sun}}\ and\ \bibinfo {author} {\bibfnamefont {W.~H.}\ \bibnamefont
  {Miller}},\ }\href {\doibase 10.1063/1.473171} {\bibfield  {journal}
  {\bibinfo  {journal} {J. Chem. Phys.}\ }\textbf {\bibinfo {volume} {106}},\
  \bibinfo {pages} {916} (\bibinfo {year} {1997})}\BibitemShut {NoStop}%
\bibitem [{\citenamefont {Sun}\ \emph {et~al.}(1998{\natexlab{a}})\citenamefont
  {Sun}, \citenamefont {Wang},\ and\ \citenamefont {Miller}}]{Sun1998}%
  \BibitemOpen
  \bibfield  {author} {\bibinfo {author} {\bibfnamefont {X.}~\bibnamefont
  {Sun}}, \bibinfo {author} {\bibfnamefont {H.}~\bibnamefont {Wang}}, \ and\
  \bibinfo {author} {\bibfnamefont {W.~H.}\ \bibnamefont {Miller}},\ }\href
  {\doibase 10.1063/1.477025} {\bibfield  {journal} {\bibinfo  {journal} {J.
  Chem. Phys.}\ }\textbf {\bibinfo {volume} {109}},\ \bibinfo {pages} {4190}
  (\bibinfo {year} {1998}{\natexlab{a}})}\BibitemShut {NoStop}%
\bibitem [{\citenamefont {Sun}\ \emph {et~al.}(1998{\natexlab{b}})\citenamefont
  {Sun}, \citenamefont {Wang},\ and\ \citenamefont {Miller}}]{Sun1998-1}%
  \BibitemOpen
  \bibfield  {author} {\bibinfo {author} {\bibfnamefont {X.}~\bibnamefont
  {Sun}}, \bibinfo {author} {\bibfnamefont {H.}~\bibnamefont {Wang}}, \ and\
  \bibinfo {author} {\bibfnamefont {W.~H.}\ \bibnamefont {Miller}},\ }\href
  {\doibase 10.1063/1.477389} {\bibfield  {journal} {\bibinfo  {journal} {J.
  Chem. Phys.}\ }\textbf {\bibinfo {volume} {109}},\ \bibinfo {pages} {7064}
  (\bibinfo {year} {1998}{\natexlab{b}})}\BibitemShut {NoStop}%
\bibitem [{\citenamefont {Miller}(1970)}]{Miller_S-Matrix_1970}%
  \BibitemOpen
  \bibfield  {author} {\bibinfo {author} {\bibfnamefont {W.~H.}\ \bibnamefont
  {Miller}},\ }\href {\doibase 10.1063/1.1674535} {\bibfield  {journal}
  {\bibinfo  {journal} {J. Chem. Phys.}\ }\textbf {\bibinfo {volume} {53}},\
  \bibinfo {pages} {3578} (\bibinfo {year} {1970})}\BibitemShut {NoStop}%
\bibitem [{\citenamefont {Grossmann}\ and\ \citenamefont
  {Xavier}(1998)}]{Grossmann_Xavier_SCderivation_1998}%
  \BibitemOpen
  \bibfield  {author} {\bibinfo {author} {\bibfnamefont {F.}~\bibnamefont
  {Grossmann}}\ and\ \bibinfo {author} {\bibfnamefont {A.~L.}\ \bibnamefont
  {Xavier}},\ }\href@noop {} {\bibfield  {journal} {\bibinfo  {journal} {Phys.
  Lett. A}\ }\textbf {\bibinfo {volume} {243}},\ \bibinfo {pages} {243}
  (\bibinfo {year} {1998})}\BibitemShut {NoStop}%
\bibitem [{\citenamefont {Buchholz}\ \emph {et~al.}(2016)\citenamefont
  {Buchholz}, \citenamefont {Grossmann},\ and\ \citenamefont
  {Ceotto}}]{Buchholz_Ceotto_MixedSC_2016}%
  \BibitemOpen
  \bibfield  {author} {\bibinfo {author} {\bibfnamefont {M.}~\bibnamefont
  {Buchholz}}, \bibinfo {author} {\bibfnamefont {F.}~\bibnamefont {Grossmann}},
  \ and\ \bibinfo {author} {\bibfnamefont {M.}~\bibnamefont {Ceotto}},\
  }\href@noop {} {\bibfield  {journal} {\bibinfo  {journal} {J. Chem. Phys.}\
  }\textbf {\bibinfo {volume} {144}},\ \bibinfo {pages} {094102} (\bibinfo
  {year} {2016})}\BibitemShut {NoStop}%
\bibitem [{\citenamefont {Grossmann}(2006)}]{Grossmann_SChybrid_2006}%
  \BibitemOpen
  \bibfield  {author} {\bibinfo {author} {\bibfnamefont {F.}~\bibnamefont
  {Grossmann}},\ }\href
  {http://scitation.aip.org/content/aip/journal/jcp/125/1/10.1063/1.2213255}
  {\bibfield  {journal} {\bibinfo  {journal} {J. Chem. Phys.}\ }\textbf
  {\bibinfo {volume} {125}},\ \bibinfo {pages} {014111} (\bibinfo {year}
  {2006})}\BibitemShut {NoStop}%
\bibitem [{\citenamefont {Ma}\ and\ \citenamefont
  {Coker}(2008)}]{Coker_2008I2Kr}%
  \BibitemOpen
  \bibfield  {author} {\bibinfo {author} {\bibfnamefont {Z.}~\bibnamefont
  {Ma}}\ and\ \bibinfo {author} {\bibfnamefont {D.}~\bibnamefont {Coker}},\
  }\href@noop {} {\bibfield  {journal} {\bibinfo  {journal} {J. Chem. Phys.}\
  }\textbf {\bibinfo {volume} {128}},\ \bibinfo {pages} {244108} (\bibinfo
  {year} {2008})}\BibitemShut {NoStop}%
\bibitem [{\citenamefont {Ovchinnikov}\ and\ \citenamefont
  {Apkarian}(1996)}]{Ovchinnikov_Apkarian_Condensedphase_1996}%
  \BibitemOpen
  \bibfield  {author} {\bibinfo {author} {\bibfnamefont {M.}~\bibnamefont
  {Ovchinnikov}}\ and\ \bibinfo {author} {\bibfnamefont {V.}~\bibnamefont
  {Apkarian}},\ }\href@noop {} {\bibfield  {journal} {\bibinfo  {journal} {J.
  Chem. Phys.}\ }\textbf {\bibinfo {volume} {105}},\ \bibinfo {pages} {10312}
  (\bibinfo {year} {1996})}\BibitemShut {NoStop}%
\bibitem [{\citenamefont {Ovchinnikov}\ \emph {et~al.}(1997)\citenamefont
  {Ovchinnikov}, \citenamefont {Apkarian} \emph
  {et~al.}}]{Ovchinnikov_Apkarian_Ramanspectra_1997}%
  \BibitemOpen
  \bibfield  {author} {\bibinfo {author} {\bibfnamefont {M.}~\bibnamefont
  {Ovchinnikov}}, \bibinfo {author} {\bibfnamefont {V.}~\bibnamefont
  {Apkarian}},  \emph {et~al.},\ }\href@noop {} {\bibfield  {journal} {\bibinfo
   {journal} {J. Chem. Phys.}\ }\textbf {\bibinfo {volume} {106}},\ \bibinfo
  {pages} {5775} (\bibinfo {year} {1997})}\BibitemShut {NoStop}%
\bibitem [{\citenamefont {Kaledin}\ and\ \citenamefont
  {Miller}(2003{\natexlab{a}})}]{Kaledin_Miller_Timeaveraging_2003}%
  \BibitemOpen
  \bibfield  {author} {\bibinfo {author} {\bibfnamefont {A.~L.}\ \bibnamefont
  {Kaledin}}\ and\ \bibinfo {author} {\bibfnamefont {W.~H.}\ \bibnamefont
  {Miller}},\ }\href {\doibase 10.1063/1.1562158} {\bibfield  {journal}
  {\bibinfo  {journal} {J. Chem. Phys.}\ }\textbf {\bibinfo {volume} {118}},\
  \bibinfo {pages} {7174} (\bibinfo {year} {2003}{\natexlab{a}})}\BibitemShut
  {NoStop}%
\bibitem [{\citenamefont {Elran}\ and\ \citenamefont
  {Kay}(1999)}]{Elran_Kay_ImprovingHK_1999}%
  \BibitemOpen
  \bibfield  {author} {\bibinfo {author} {\bibfnamefont {Y.}~\bibnamefont
  {Elran}}\ and\ \bibinfo {author} {\bibfnamefont {K.}~\bibnamefont {Kay}},\
  }\href@noop {} {\bibfield  {journal} {\bibinfo  {journal} {J. Chem. Phys.}\
  }\textbf {\bibinfo {volume} {110}},\ \bibinfo {pages} {3653} (\bibinfo {year}
  {1999})}\BibitemShut {NoStop}%
\bibitem [{\citenamefont {Kaledin}\ and\ \citenamefont
  {Miller}(2003{\natexlab{b}})}]{Kaledin_Miller_TAmolecules_2003}%
  \BibitemOpen
  \bibfield  {author} {\bibinfo {author} {\bibfnamefont {A.~L.}\ \bibnamefont
  {Kaledin}}\ and\ \bibinfo {author} {\bibfnamefont {W.~H.}\ \bibnamefont
  {Miller}},\ }\href {\doibase 10.1063/1.1589477} {\bibfield  {journal}
  {\bibinfo  {journal} {J. Chem. Phys.}\ }\textbf {\bibinfo {volume} {119}},\
  \bibinfo {pages} {3078} (\bibinfo {year} {2003}{\natexlab{b}})}\BibitemShut
  {NoStop}%
\bibitem [{\citenamefont {{Di Liberto}}\ and\ \citenamefont
  {Ceotto}(2016)}]{DiLiberto_Ceotto_Prefactors_2016}%
  \BibitemOpen
  \bibfield  {author} {\bibinfo {author} {\bibfnamefont {G.}~\bibnamefont {{Di
  Liberto}}}\ and\ \bibinfo {author} {\bibfnamefont {M.}~\bibnamefont
  {Ceotto}},\ }\href@noop {} {\bibfield  {journal} {\bibinfo  {journal} {J.
  Chem. Phys.}\ }\textbf {\bibinfo {volume} {145}},\ \bibinfo {pages} {144107}
  (\bibinfo {year} {2016})}\BibitemShut {NoStop}%
\bibitem [{\citenamefont {Goletz}\ and\ \citenamefont
  {Grossmann}(2009)}]{Goletz_Grossmann_Decoherence_2009}%
  \BibitemOpen
  \bibfield  {author} {\bibinfo {author} {\bibfnamefont {C.-M.}\ \bibnamefont
  {Goletz}}\ and\ \bibinfo {author} {\bibfnamefont {F.}~\bibnamefont
  {Grossmann}},\ }\href@noop {} {\bibfield  {journal} {\bibinfo  {journal} {J.
  Chem. Phys.}\ }\textbf {\bibinfo {volume} {130}},\ \bibinfo {pages} {244107}
  (\bibinfo {year} {2009})}\BibitemShut {NoStop}%
\bibitem [{\citenamefont {Goletz}\ \emph {et~al.}(2010)\citenamefont {Goletz},
  \citenamefont {Koch},\ and\ \citenamefont {Grossmann}}]{Goletz2010}%
  \BibitemOpen
  \bibfield  {author} {\bibinfo {author} {\bibfnamefont {C.-M.}\ \bibnamefont
  {Goletz}}, \bibinfo {author} {\bibfnamefont {W.}~\bibnamefont {Koch}}, \ and\
  \bibinfo {author} {\bibfnamefont {F.}~\bibnamefont {Grossmann}},\ }\href
  {\doibase 10.1016/j.chemphys.2010.06.019} {\bibfield  {journal} {\bibinfo
  {journal} {Chem. Phys.}\ }\textbf {\bibinfo {volume} {375}},\ \bibinfo
  {pages} {227} (\bibinfo {year} {2010})}\BibitemShut {NoStop}%
\bibitem [{\citenamefont {Tao}\ and\ \citenamefont
  {Miller}(2009{\natexlab{a}})}]{Tao_Miller_I2Ar_2009_1}%
  \BibitemOpen
  \bibfield  {author} {\bibinfo {author} {\bibfnamefont {G.}~\bibnamefont
  {Tao}}\ and\ \bibinfo {author} {\bibfnamefont {W.~H.}\ \bibnamefont
  {Miller}},\ }\href {\doibase 10.1063/1.3132224} {\bibfield  {journal}
  {\bibinfo  {journal} {The Journal of Chemical Physics}\ }\textbf {\bibinfo
  {volume} {130}},\ \bibinfo {pages} {184108} (\bibinfo {year}
  {2009}{\natexlab{a}})}\BibitemShut {NoStop}%
\bibitem [{\citenamefont {Tao}\ and\ \citenamefont
  {Miller}(2009{\natexlab{b}})}]{Tao_Miller_I2Ar_2009_2}%
  \BibitemOpen
  \bibfield  {author} {\bibinfo {author} {\bibfnamefont {G.}~\bibnamefont
  {Tao}}\ and\ \bibinfo {author} {\bibfnamefont {W.~H.}\ \bibnamefont
  {Miller}},\ }\href {\doibase 10.1063/1.3271241} {\bibfield  {journal}
  {\bibinfo  {journal} {The Journal of Chemical Physics}\ }\textbf {\bibinfo
  {volume} {131}},\ \bibinfo {pages} {224107} (\bibinfo {year}
  {2009}{\natexlab{b}})}\BibitemShut {NoStop}%
\bibitem [{\citenamefont {Pan}\ and\ \citenamefont
  {Tao}(2013)}]{Tao_Feng_2013}%
  \BibitemOpen
  \bibfield  {author} {\bibinfo {author} {\bibfnamefont {F.}~\bibnamefont
  {Pan}}\ and\ \bibinfo {author} {\bibfnamefont {G.}~\bibnamefont {Tao}},\
  }\href {\doibase 10.1063/1.4794191} {\bibfield  {journal} {\bibinfo
  {journal} {The Journal of Chemical Physics}\ }\textbf {\bibinfo {volume}
  {138}},\ \bibinfo {pages} {091101} (\bibinfo {year} {2013})}\BibitemShut
  {NoStop}%
\bibitem [{\citenamefont {Buchholz}\ \emph {et~al.}(2012)\citenamefont
  {Buchholz}, \citenamefont {Goletz}, \citenamefont {Grossmann}, \citenamefont
  {Schmidt}, \citenamefont {Heyda},\ and\ \citenamefont
  {Jungwirth}}]{Buchholz_Jungwirth_CondensedPhase_2012}%
  \BibitemOpen
  \bibfield  {author} {\bibinfo {author} {\bibfnamefont {M.}~\bibnamefont
  {Buchholz}}, \bibinfo {author} {\bibfnamefont {C.-M.}\ \bibnamefont
  {Goletz}}, \bibinfo {author} {\bibfnamefont {F.}~\bibnamefont {Grossmann}},
  \bibinfo {author} {\bibfnamefont {B.}~\bibnamefont {Schmidt}}, \bibinfo
  {author} {\bibfnamefont {J.}~\bibnamefont {Heyda}}, \ and\ \bibinfo {author}
  {\bibfnamefont {P.}~\bibnamefont {Jungwirth}},\ }\href@noop {} {\bibfield
  {journal} {\bibinfo  {journal} {J. Phys. Chem. A}\ }\textbf {\bibinfo
  {volume} {116}},\ \bibinfo {pages} {11199} (\bibinfo {year}
  {2012})}\BibitemShut {NoStop}%
\bibitem [{\citenamefont {Bihary}\ \emph {et~al.}(2001)\citenamefont {Bihary},
  \citenamefont {Gerber},\ and\ \citenamefont
  {Apkarian}}]{Bihary_Apkarian_VSCF_2001}%
  \BibitemOpen
  \bibfield  {author} {\bibinfo {author} {\bibfnamefont {Z.}~\bibnamefont
  {Bihary}}, \bibinfo {author} {\bibfnamefont {R.}~\bibnamefont {Gerber}}, \
  and\ \bibinfo {author} {\bibfnamefont {V.}~\bibnamefont {Apkarian}},\
  }\href@noop {} {\bibfield  {journal} {\bibinfo  {journal} {J. Chem. Phys.}\
  }\textbf {\bibinfo {volume} {115}},\ \bibinfo {pages} {2695} (\bibinfo {year}
  {2001})}\BibitemShut {NoStop}%
\bibitem [{\citenamefont {Karavitis}\ \emph {et~al.}(2005)\citenamefont
  {Karavitis}, \citenamefont {Kumada}, \citenamefont {Goldschleger},\ and\
  \citenamefont {Apkarian}}]{Karavitis2005}%
  \BibitemOpen
  \bibfield  {author} {\bibinfo {author} {\bibfnamefont {M.}~\bibnamefont
  {Karavitis}}, \bibinfo {author} {\bibfnamefont {T.}~\bibnamefont {Kumada}},
  \bibinfo {author} {\bibfnamefont {I.~U.}\ \bibnamefont {Goldschleger}}, \
  and\ \bibinfo {author} {\bibfnamefont {V.~A.}\ \bibnamefont {Apkarian}},\
  }\href {\doibase 10.1039/B416143B} {\bibfield  {journal} {\bibinfo  {journal}
  {Phys. Chem. Chem. Phys.}\ }\textbf {\bibinfo {volume} {7}},\ \bibinfo
  {pages} {791} (\bibinfo {year} {2005})}\BibitemShut {NoStop}%
\bibitem [{\citenamefont {R{\"o}blitz}\ \emph {et~al.}(2014)\citenamefont
  {R{\"o}blitz}, \citenamefont {Schmidt},\ and\ \citenamefont
  {Weber}}]{Schmidt2014}%
  \BibitemOpen
  \bibfield  {author} {\bibinfo {author} {\bibfnamefont {S.}~\bibnamefont
  {R{\"o}blitz}}, \bibinfo {author} {\bibfnamefont {B.}~\bibnamefont
  {Schmidt}}, \ and\ \bibinfo {author} {\bibfnamefont {M.}~\bibnamefont
  {Weber}},\ }\href {https://sourceforge.net/projects/trajlab} {\enquote
  {\bibinfo {title} {{TrajLab: MATLAB based programs for trajectory simulations
  of molecules}},}\ } (\bibinfo {year} {2014})\BibitemShut {NoStop}%
\bibitem [{\citenamefont {Zadoyan}\ \emph {et~al.}(1994)\citenamefont
  {Zadoyan}, \citenamefont {Li}, \citenamefont {Martens},\ and\ \citenamefont
  {Apkarian}}]{Zadoya1994}%
  \BibitemOpen
  \bibfield  {author} {\bibinfo {author} {\bibfnamefont {R.}~\bibnamefont
  {Zadoyan}}, \bibinfo {author} {\bibfnamefont {Z.}~\bibnamefont {Li}},
  \bibinfo {author} {\bibfnamefont {C.~C.}\ \bibnamefont {Martens}}, \ and\
  \bibinfo {author} {\bibfnamefont {V.~A.}\ \bibnamefont {Apkarian}},\ }\href
  {\doibase 10.1063/1.468359} {\bibfield  {journal} {\bibinfo  {journal} {J.
  Chem. Phys.}\ }\textbf {\bibinfo {volume} {101}},\ \bibinfo {pages} {6648}
  (\bibinfo {year} {1994})}\BibitemShut {NoStop}%
\bibitem [{\citenamefont {Karavitis}\ and\ \citenamefont
  {Apkarian}(2004)}]{Karavitis2004}%
  \BibitemOpen
  \bibfield  {author} {\bibinfo {author} {\bibfnamefont {M.}~\bibnamefont
  {Karavitis}}\ and\ \bibinfo {author} {\bibfnamefont {V.~A.}\ \bibnamefont
  {Apkarian}},\ }\href {\doibase 10.1063/1.1630567} {\bibfield  {journal}
  {\bibinfo  {journal} {J. Chem. Phys.}\ }\textbf {\bibinfo {volume} {120}},\
  \bibinfo {pages} {292} (\bibinfo {year} {2004})}\BibitemShut {NoStop}%
\bibitem [{\citenamefont {Liberto}\ \emph {et~al.}(2018)\citenamefont
  {Liberto}, \citenamefont {Conte},\ and\ \citenamefont
  {Ceotto}}]{DiLiberto_Ceotto_Jacobiano_2018}%
  \BibitemOpen
  \bibfield  {author} {\bibinfo {author} {\bibfnamefont {G.~D.}\ \bibnamefont
  {Liberto}}, \bibinfo {author} {\bibfnamefont {R.}~\bibnamefont {Conte}}, \
  and\ \bibinfo {author} {\bibfnamefont {M.}~\bibnamefont {Ceotto}},\
  }\href@noop {} {\bibfield  {journal} {\bibinfo  {journal} {J. Chem. Phys.}\
  }\textbf {\bibinfo {volume} {148}},\ \bibinfo {pages} {014307} (\bibinfo
  {year} {2018})}\BibitemShut {NoStop}%
\end{thebibliography}%

\end{document}